\documentclass[11pt]{article}

\pdfoutput=1

\usepackage[utf8]{inputenc}
\usepackage{jcappub}
\usepackage[export]{adjustbox}
\usepackage[english]{babel}
\usepackage{amsmath}
\usepackage{color}
\usepackage{epstopdf}
\usepackage{graphicx}
\usepackage{braket}
\usepackage{axodraw4j}
\usepackage{amsmath}
\usepackage{amsmath}
\usepackage{slashed}
\usepackage{amssymb}
\usepackage{braket}
\usepackage{amsmath}
\usepackage{braket}
\usepackage{epigraph}
\usepackage{tensor}
\usepackage{enumerate}
\usepackage{subfig}
\usepackage{verbatim}
\usepackage{hyperref}

\usepackage{mathtools}
\usepackage{tikz}
\usetikzlibrary{matrix,arrows,decorations.pathmorphing,decorations.pathreplacing}

\usepackage{tikz}
\usetikzlibrary{matrix,arrows,decorations.pathmorphing,decorations.pathreplacing}
\usepackage{amsmath,amssymb,euscript,array,mathrsfs,appendix,ctable,marvosym}
\usepackage{arydshln}
\usepackage{todonotes}
\usepackage{tikz}
\usepackage{pgfplots}
\pgfplotsset{/pgf/number format/use comma,compat=newest}
\usepackage{color}
\usepackage{graphicx}  

\author[a]{Francesca V. Day}
\author[b]{and Jamie I. McDonald}

\affiliation[a]{DAMTP, Centre for Mathematical Sciences,\\
	 Wilberforce Road, Cambridge, \\
	 CB3 0WA, \\
	 United Kingdom}
\affiliation[b]{Technische Universität München,\newline Physik-Department, James-Franck-Straße, \newline
	85748 Garching, \newline
	Germany}

\emailAdd{francesca.day@maths.cam.ac.uk}
\emailAdd{jamie.mcdonald@tum.de}

	\abstract{It is a well-known fact that compact gravitating objects admit bound state configurations for massive bosonic fields. In this work we describe a new class of superradiant instabilities of axion bound states in neutron star magnetospheres. The instability arises from the mixing of axion and photon modes in the magnetic field of the neutron star which extract energy from the rotating magnetosphere. Unlike for black holes, where the dissipation required for superradiance is provided by an absorptive horizon, the non-hermitian dynamics in this paper come from the resistivity in the stellar magnetosphere arising from a finite bulk conductivity. The axion field mixes with photon modes which superradiantly scatter off the magnetosphere, extracting rotational energy which is then deposited back into the axion sector leading to an instability. We derive the superradiant eigenfrequencies for the axion-photon system using quantum mechanical perturbation theory on the axion boundstate, drawing an analogy with atomic selection rules. We then compare the characteristic time scale of the instability to the spin-down measurements of pulsars which limit the allowed rate of angular momentum extraction from neutron stars.  }

\title{Axion superradiance in rotating neutron stars}
\notoc

\begin{document}
	\maketitle
	\newpage
	\tableofcontents

	\section{Introduction}
	
	The stability of compact astrophysical objects to perturbations has been the subject of many studies spanning several decades, focussing both on black hole and stellar environments. In the context of black holes, this involves their so-called \textit{quasi-normal-modes} \cite{Kokkotas:1999bd, Konoplya:2011qq, Pani:2013pma} with similar efforts also carried out in the context of neutron stars \cite{Andersson:2000mf}. Through this work, a rich class of perturbations and their instabilities has emerged, which are of interest not only as phenomena in their own right, but also as tools to constrain new physics through astrophysical observations.

	 Of particular interest is the study of ultra-light bosonic fields \cite{Svrcek:2006yi,Arvanitaki:2009fg} which can develop macroscopic configurations around astrophysical objects when their Compton wavelength is sufficiently large. One of the most well-known phenomena occurs for rotating black holes, in which ultra light degrees of freedom with masses below the frequency of the black hole extract rotational energy via the dissipative dynamics provided by a horizon, which acts as a one-way membrane. This is the phenomenon of black hole superradiance \cite{Arvanitaki:2010sy,Pani:2012vp,Arvanitaki:2014wva,Brito:2014wla,Brito:2015oca,Rosa:2017ury,Filippini:2019cqk}. Similar mechanisms can also occur in stars \cite{Cardoso:2015zqa,Richartz:2013hza,Cardoso:2017kgn}, with some other form of dissipative dynamics in in lieu of the horizon. It is the latter case which will be the subject of the present work.

	 From the no-hair theorem, the sparsity of parameters associated to idealised black holes -- e.g. charge, mass and angular momentum -- makes an analysis of their associated instabilities relatively clean from a theoretical point of view, with the growth rate of the instabilities of an ideal Kerr black hole determined essentially by the the mass $\mu$ of the bosonic field and the mass $M$ and angular velocity $\Omega$ of the black hole \cite{Brito:2015oca}. Stellar environments by contrast, posses a richer class of instabilities and parameters associated to the plasma. However, pulsars in particular have one important observational advantage over black holes, namely that there is an extensive catalogue of pulsar measurements \cite{ATNF} currently unrivalled both in extent and precision in comparison to observations of black holes, which by their very nature are inherently more illusive. For instance, whilst we have yet to accurately measure the spin-down rate of black holes, the spin and the  spindown  rate  of  pulsars  are constrained with great precision through pulsar timing \cite{Lorimer}. Such sensitivity is vitally important if one hopes to constrain new physics via pulsar observations, either through the rate of angular momentum extraction from ultra-light fields \cite{Cardoso:2017kgn} or polarisation-dependent effects on radio-emissions passing through scalar configurations around astrophysical objects \cite{Plascencia:2017kca,Mohanty:1993nh}.

	  Ultra-light bosonic fields occur in many well-motivated extensions of the Standard Model. In particular, axions are an attractive dark matter candidate \cite{1510.07633}, as well offering a solution to the strong CP problem \cite{Peccei:1977hh,Weinberg:1977ma,Wilczek:1977pj} and arising generically in string compactifications \cite{hep-th/0602233,hep-th/0605206}. Substantial experimental and theoretical effort has already been made in constraining the axion parameter space, in particular placing bounds on the axion-photon coupling $g_{a \gamma \gamma}$ \cite{Patrignani:2016xqp} considered in this work. For low mass ($\mu \lesssim 10^{-12} \, {\rm eV}$) axions, the leading bounds on the axion photon coupling come from consideration of axion production in the nearby supernova SN 1987A \cite{Payez:2014xsa}, and from non-observation of axion-photon mixing in the magnetic fields of galaxy clusters \cite{Berg:2016ese, Marsh:2017yvc,Conlon:2017qcw}. These constrain $g_{a \gamma \gamma} \lesssim 10^{-12} \, {\rm GeV}^{-1}$. Here we present a novel phenomenological effect of the axion-photon interaction. We do not consider other interactions of the axion with the Standard Model, and treat the axion's mass and coupling as independent parameters. Our work is therefore most relevant to general string axions and need not apply specifically to the QCD axion. Furthermore, we remain agnostic as to whether axions make up any of the dark matter density.

	 	In this paper we describe a new class of instabilities arising from axion-photon mixing in the magnetic fields and plasma of neutron stars. Specifically, we show how the presence of dissipative dynamics in the neutron star magnetosphere leads to the extraction of rotational energy by macroscopic axion-photon solutions, leading to a superradiant growth of the axion profile. Before proceeding to the main bulk of the calculation, we take the opportunity now to sketch the physical nature of the mechanism, which is a multi-step process consisting of the following stages. \\
	 	
	 	\newpage

	 	\textbf{The Mechanism}
	 	\begin{enumerate}[(i)]
	 		\item Neutron stars, like black holes, have a confining exterior Schwarzschild potential  giving rise to hydrogen-like bound state solutions (see sec.~\ref{Boundstates}) for massive fields. The system is therefore sometimes referred to as a \textit{gravitational-atom}, with boundstates labelled by the usual ``quantum numbers" $(\ell, m,n)$ with $\ell,m$ labelling angular momentum and $n$ the energy levels. For stationary backgrounds, the fields have harmonic time-dependence $\sim e^{-i \omega t}$ with quantitised eigenfrequencies $\omega_{\ell m n}$ associated to each boundstate. Gravitational boundstates of black holes have complex frequencies owing to the presence of a horizon which acts as a one-way membrane. By contrast, the frequencies of the stellar gravitational boundstates (in the absence of any further dissipative mechanism) are real \cite{Plascencia:2017kca}.  One can now imagine perturbing the neutron star with one such a boundstate. This could arise, most conservatively, for example, via an initial quantum fluctuation in the axion field. Since we deal with neutron stars the boundstate could also arise from background electromagnetic fields of the neutron star which source the axion via $F^{\mu \nu}\tilde{F}_{\mu \nu}=-4 \textbf{E}\cdot \textbf{B}$, as shown in \cite{Garbrecht:2018akc}.
	 		\item  Next, the axion field couples to a photon mode via the axion-photon mixing induced by the background magnetic field of the neutron star. These photon modes then interact with the magnetosphere and, due to non-hermitian dynamics, are amplified as they scatter, extracting rotational energy from the background plasma. In the present context the dissipation is provided by a finite bulk conductivity $\sigma$, and an azimuthal fluid velocity $u_\varphi$ provides a source of rotational energy. For a neutron star rotating sufficiently quickly, this dissipative interaction between the photon and the neutron star magnetosphere leads to superradiant amplification of the photon modes.
	 		\item The photon mode then transfers this energy gain back into the axion sector, where it is trapped due to the confining nature of the axion potential, causing the axion boundstate to grow. This process repeats ad infinitum, leading to an instability and exponential growth of the axion profile with time. 
	 	\end{enumerate}

	 	\begin{figure}
	 		\centering
	 		\includegraphics[scale=0.6]{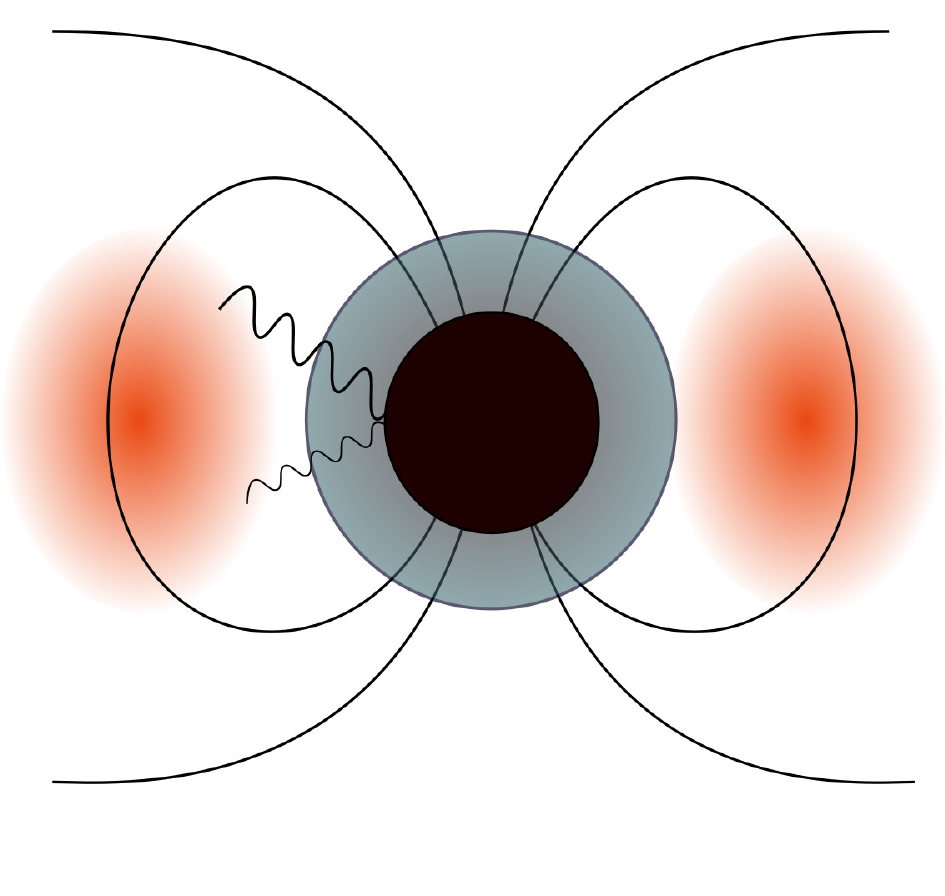}
	 		\caption{Schematic illustration of the instability. The axion boundstate (orange) mixes with a photon mode which is then amplified by scattering off the rotating magnetosphere (grey). The photon energy is then deposited back into the axion sector. }
	 	\end{figure}
	 	
	 	More properly one can view the problem in terms of flavour versus mass-eigenstates. Since the axion and photon modes mix in the magnetic field of the neutron star, the physically relevant mass-eigenstates will be a superposition of photon and axion ``flavour" states. The heavier of the mass-eigenstates has a mass of order the axion mass $\sim \mu$ (plus small corrections arising from the mixing) and admits gravitational boundstates. Since this heavier mass boundstate contains a photon component it inherits some of the non-hermitian dynamics associated to the photon/plasma sector leading it to grow and trigger an instability.

	 	One should remark that setup (ii) is reminiscent of the superradiant scattering of dark photons off a conducting star considered in \cite{Cardoso:2017kgn} with two important differences. The first is that in that instance, the dark photon field, $A^\mu$, was given a finite mass, such that the field $A^{\mu}$ itself admitted superradiant bound-states. By contrast, the massless photon field in our setup plays merely an intermediary role. It is able to extract rotational energy and superradiantly scatter, but this gain in energy would otherwise be lost as the photon escapes to infinity were it not for the fact it is recaptured by the axion profile. The second important difference, is that our our macroscopic bulk conductivity $\sigma$ arises from the magnetosphere rather than the neutron star. It is therefore typically much lower than that encountered in the star itself. This point is discussed more in sec.~\ref{Conductvity}. As a result, we avoid the large-$\sigma$ suppression which greatly decreases the amplification factor of scattered photon modes and which led the author's of \cite{Cardoso:2017kgn} to consider a lower conductivity arising from the physics  of an additional dark sector.  
	 	
	 	The structure of the paper is as follows. Sec.~\ref{theory} outlines the basic model for axion-photon mixing, and introduces dissipation via Ohm's law. In sec.~\ref{1D} we study instabilities in homogeneous plasma, which provides a useful basis for understanding the more complicated neutron star setup. In sec.~\ref{3D} we set out the problem of finding the superradiant eigenfrequencies in the context of neutron stars, which involves the mixing of the axion and photon fields in a background magnetic field in 3D. In sec.~\ref{QMPT} we show how, by analogy with standard atomic perturbation theory in quantum mechanics, one can derive the superradiant eigenfrequencies perturbatively in $\sigma$ and $g_{a \gamma \gamma}$. The superradiant corrections are calculable in a similar way to atomic transitions as dictated by selection rules. We then demonstrate in sec.~\ref{Example} an explicit calculation of the eigenfrequencies for a simple magnetic field configuration and a co-rotating magnetospheric fluid. Finally we compare the instability time scale to the characteristic age of neutron stars as given by their spin-down rates. We offer our conclusions in sec.~\ref{Discussion} and speculate on how other types of unstable modes in plasma might lead to similar effects.

	\section{Basic Theory}\label{theory}
	
	The Lagrangian for the axion and photon fields is:
	
	\begin{equation} 
	\mathcal{L} \supset \sqrt{-g} \left[ \frac{1}{2} \partial_{\mu} \phi \partial^{\mu} \phi - \frac{\mu^2}{2} \phi^2 - \frac{1}{4} F_{\mu \nu} F^{\mu \nu} - \frac{g_{a \gamma \gamma}}{4} \phi F_{\mu \nu} \tilde{F}^{\mu \nu} - A_{\mu} j^{\mu} \right],
	\end{equation}
	where $\tilde{F}^{\mu \nu} = \epsilon^{\mu \nu \gamma \delta}/2\, F_{\gamma \delta}$, $g_{a \gamma \gamma}$ is the axion-photon coupling and $j^{\mu}$ is the current in the surrounding medium. We also have the covariant form of Ohm's law:
	
	\begin{equation}
	j^\mu = \sigma  F^{\mu \nu} u_\nu + \rho u^\mu, \qquad \quad  u^\mu u_\mu = 1,
	\end{equation}
	where $\rho$, $\sigma$ and $u$ are the electric charge, conductivity and four-velocity of the medium, respectively. This leads to the equations of motion:

	\begin{equation}
	\square \phi + \mu^2 \phi = -\frac{ g_{a \gamma \gamma}}{4}F_{\mu \nu}\tilde{F}^{\mu \nu},  \label{EOM1}
	\end{equation}
	
	\begin{equation}
	\partial_{\mu} F^{\mu \nu}  - \sigma F^{\nu \mu}u_\mu  = - g_{a \gamma \gamma} (\partial_\mu \phi) \tilde{F}^{\mu \nu} + \rho u^\mu.  \label{EOM2}
	\end{equation}
	Consider linearised axion-photon fluctuations $\phi$ and $A^\mu$ about a background electromagnetic field with field strength $F_B^{\mu \nu}$. This leads to the following linearised field equations:
	\begin{equation}
	\square \phi + \mu^2 \phi = -\frac{ g_{a \gamma \gamma}}{2}F_{\mu \nu}\tilde{F}^{\mu \nu}_B,  \label{EOM3}
	\end{equation}
	\begin{equation}
	\partial_{\mu} F^{\mu \nu}  - \sigma F^{\nu \mu}u_\mu  = - g_{a \gamma \gamma} (\partial_\mu \phi) \tilde{F}^{\mu \nu}_B , \label{EOM4}
	\end{equation}
	where $F^{\mu \nu}$ is the field strength of the fluctuation field. Note that we have not included fluctuations in the fluid velocity and density. As the fluid does not couple directly to the axion in our system, these fluctuations would contribute only at a higher order in perturbation theory to our calculations in section \ref{3D}.
	
	
	\section{Axion-photon instabilities in homogeneous plasmas}\label{1D}
	Consider now the case of constant background fields, $F_{B}^{\mu \nu}$, $u^\mu$ and $\sigma$. In this case one can immediately Fourier transform the equations of motion \eqref{EOM3}-\eqref{EOM4}, which in Lorenz gauge $\partial_\mu A^\mu =0$, read
	\begin{equation}
	\left(
	\begin{array}{cc}
	-k^2 + \mu^2  & \quad -i g_{a \gamma \gamma} k_\mu \tilde{F}^{\mu \nu}_B \\
-	i g_{a \gamma \gamma} k_\nu \tilde{F}^{\nu \mu}_B &\quad  -k^2 \eta^{\mu \nu} +i \sigma ( k^\mu u^\nu      - (k\cdot u)  \eta^{\mu \nu})
	\end{array}
	\right)
	\left(
	\begin{array}{c}
	\phi(k)\\
	A_\nu(k)
	\end{array}
	\right) = 0.\label{FourierMatrix}
	\end{equation}
	The vanishing of the matrix determinant in (\ref{FourierMatrix}) gives the following dispersion relations
	\begin{equation}
	k^2 \Big[ k^2 + i \sigma (k \cdot u) \Big]^2 \Big[ g_{a \gamma \gamma}^2 (k \cdot \tilde{F}^2_B \cdot k) - (k^2 - \mu^2)(k^2 + i \sigma k \cdot u)   \Big] =0. \label{Determinant}
	\end{equation}
	The second term clearly describes the effect of plasma resistivity on photon fluctuations. The third term relates to the effects of axion-photon mixing and is similar to the dispersion relation found in \cite{1811.04945}, with the additional feature that we have now a fluid velocity $u^\mu$.  Explicitly, in terms of the electric and magnetic fields $\textbf{E}$ and $\textbf{B}$ contained in the background field strength $F_B^{\mu \nu}$, one finds the following dispersion relation:
	\begin{equation}
	\omega^2 - \textbf{k}^2 - \mu^2  = g^2_{a \gamma \gamma}\left[ \frac{ \omega^2 |\textbf{B}|^2-(\textbf{k} \cdot \textbf{B})^2  - 2\textbf{k}\cdot (\textbf{E}\times \textbf{B}) \omega  + |\textbf{k}|^2 |\textbf{E}|^2 -(\textbf{E}\cdot \textbf{k})^2}{\omega^2 - \textbf{k}^2 + i \sigma (k \cdot u) }\right] . \label{dispersion1}
	\end{equation}

	\subsection{An aside: instabilities from electromagnetic fields} 

	We deal first with the case $\sigma =0$, $u=0$ in the absence of plasma, and ask what, if any, are the instabilities that might arise from background electromagnetic fields alone. This reduces to the question of whether or not the quartic polynomial in $\omega$ of eq.~(\ref{dispersion1}) has a complex root, i.e. whether there are frequencies with a non-vanishing imaginary part $\text{Im}(\omega) \neq0$. This is determined by the sign of the discriminant of the quartic. 	For the special case $\textbf{B} =0$, there is an instability whenever
	\begin{equation}
\textbf{k}^2+ \mu^2 < \sin^2 \theta_E \,g^2_{a \gamma \gamma}\textbf{E}^2,
	\end{equation}
	where $\theta_E$ is the angle between $\textbf{E}$ and $\textbf{k}$. In other words, whenever the electric field or axion-photon coupling exceed some threshold value as found in \cite{1811.04945}. Meanwhile, for $\textbf{E}=0$ and $u=\sigma=0$, there is never an instability, and the eigenfrequencies $\omega$ are always real.  More generally the sign of the discriminant has a non-trivial dependence on $\mu$, $g_{a \gamma \gamma}$, $\textbf{k}$ and the electromagnetic fields. However, the plot in fig.~\ref{Instability} allows one to interpolate between these two special cases and see for what regions of parameter space $\omega$ can develop an imaginary part. 
	\pgfdeclareimage[interpolate=true]{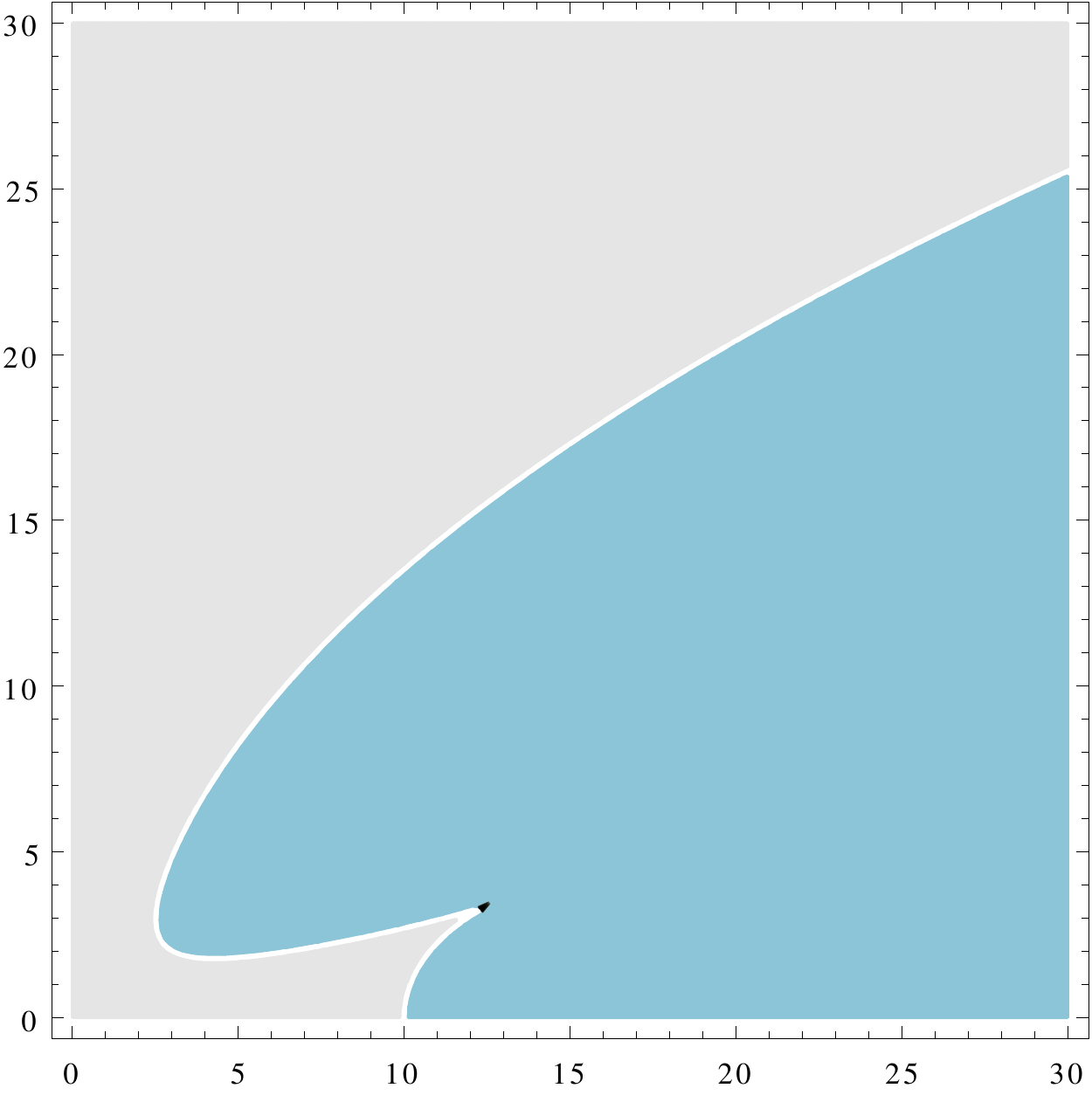}{EBEigenvalue.pdf}
	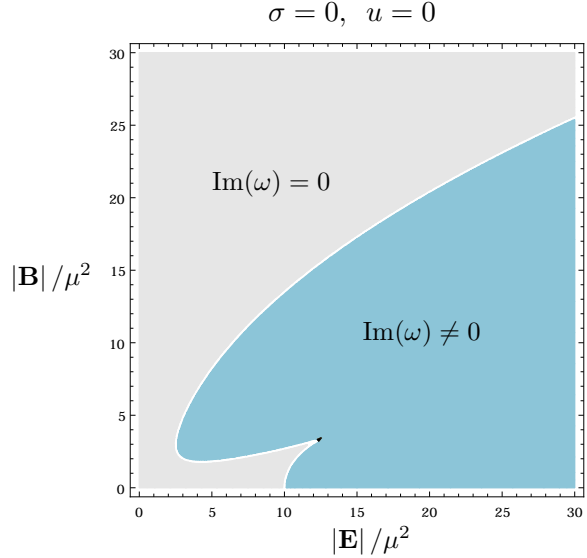
\begin{figure}[ht]
		\begin{center}
			\begin{tikzpicture}[scale=0.5]
			\pgftext[at=\pgfpoint{0cm}{0cm},left,base]{\pgfuseimage{EBEigenvalue.pdf}}
			\node at (7,-0.5) {\small  $\left|\textbf{E}\right|/\mu^2$};
			\node at (6.5,13.5) { $\sigma=0, \, \,\, u=0$};
				\node at (-1.5,6.5) {\small $\left|\textbf{B}\right|/\mu^2$};
				\node at (8,5) {\small $\quad \text{Im}(\omega) \neq 0$};
				\node at (4,9) {\small $\quad \text{Im}(\omega) = 0$};
				\node at (-1.5,9) {\small };
			\end{tikzpicture}
				\caption{Stability regions in the absence of plasma $\sigma=0$, $u=0$. The regions correspond to the sign of the discriminant associated to the quartic in eq.~(\ref{dispersion1}). We took $g_{a \gamma \gamma} \mu = 0.2$ and $k=\mu$ and the angles between $\textbf{k}$ and $\textbf{E}$, $\textbf{B}$, and the Poynting vector $\textbf{P}= \textbf{E}\times \textbf{B}$ to be, $\theta_E=\theta_B=\pi/4$ and $\theta_P=\pi/2$, respectively. }
					\label{Instability}
		\end{center}
	\end{figure}

	\subsection{Instabilities from a finite plasma velocity in a magnetic field} 
	For neutron stars, typically $\textbf{B} \gg \textbf{E}$, and the axion-photon coupling is also found to be observationally small, such that the magnetic field dominates the dispersion relation, suppressing any instability which would arise from the electric field alone. In what follows, we therefore set $\textbf{E}=0$ and consider the case only of a background magnetic field, which requires a non-vanishing plasma velocity, $u$, to generate an instability, as we now explain. Setting $\textbf{E}=0$, to leading order in axion-photon coupling, from eq.~(\ref{dispersion1}) one obtains 
	\begin{align}
	&\text{Im}\left(\omega\right)  =   \frac{g_{a \gamma \gamma}^2 \left(\omega^2 \left|\textbf{B} \right|^2 - (\textbf{k} \cdot \textbf{B})^2 \right)}{2  \omega} \cdot \frac{\sigma \left(u -v_p (1+u^2)^{1/2} \cos \theta_u  \right)}{\mu^4/\omega^2 + \sigma^2 \left( \cos \theta_u (1+u^2)^{1/2} v_p - u\right)^2  }, \nonumber \\ 
	\nonumber \\
	& v_p = \omega/k, \qquad \cos \theta_u = \hat{\textbf{k}}\cdot \hat{\textbf{u}} \label{imdeltaomega},
	\end{align}
	where $\omega$ on the right hand side should be taken as satisfying the free dispersion relation $\omega^2 = \textbf{k}^2 + \mu^2$.

	One can see that if the axion phase velocity $v_p$, is smaller than the fluid velocity then $\text{Im}(\omega)>0$ and an instability occurs. Plainly in the present setup, since $v_p \geq 1$ and $u\leq1$, the modes remain stable. This is a consequence that for linear trajectories, Lorentz invariance ensures that since $k$ and $u$ are timelike,  $k\cdot u$ must always be positive \cite{Brito:2015oca}.
	
	 However the point remains that if one can produce axions whose phase velocity is less than the fluid velocity, an instability will occur. Looking ahead to a spherical setup, consider a rotating fluid, with angular frequency $\Omega$. Since the azimuthal phase velocity of an axion mode with azimuthal number $m$ is less than the rotational velocity $\Omega$ whenever $\omega <m  \Omega$, we expect that low frequency axion modes in rotating neutron star magnetospheres will experience a superradiant instability for $\omega < m \Omega$.
	 
	  Before addressing the spherical gravitating problem in section \ref{1DFig}, it is instructive to force a velocity $v_p < u$ by introducing a Lorentz-violating sound speed $c_s<1$ for the axion such that $\square \phi \rightarrow \ddot{\phi} - c_s^2 \nabla^2 \phi$. The unperturbed axion phase velocity now reads $v_p(k)^2 = (c_s^2 + \mu^2/k^2)$ which can be inserted into (\ref{imdeltaomega}) to obtain the imaginary part of the frequency plotted in figure \ref{ImOmegaFig} for $\cos \theta_u =1$.  Assuming for simplicity ${\bf k \cdot B} = 0$, we find that superradiant amplification occurs for momenta satisfying
	  \begin{equation}
	  k^2 >k^2_c  \equiv \frac{1 + u^2}{u^2 - c_s^2 (1+u^2)} \mu^2. \label{KC} 
	  \end{equation} 
	  We also solve the resulting equations of motion mode-by-mode in a simple geometry in which $\textbf{B}$ and $\textbf{A}$ are perpendicular to fluid flow and the direction of propagation:
	\begin{equation}
	\textbf{A}\cdot \textbf{u} =0, \qquad \textbf{B} \cdot \textbf{k} = 0 \label{BGeometry} .
	\end{equation}
	In this setup, the equations for $\phi(\textbf{k},t)$ and $A_\parallel(\textbf{k},t)$ (the component of $\textbf{A}$ parallel to $\textbf{B}$) decouple, giving rise to
	\begin{align}
	&\ddot{\phi}+k^2 c_s^2 \phi + \mu^2 \phi = - g_{a \gamma \gamma} \dot{A}_\parallel B ,\nonumber\\
	&\ddot{A}_\parallel +k^2 A_\parallel+\sigma u^0\dot{A}_\parallel+  i \sigma  (\textbf{k}\cdot\textbf{u}) A_\parallel =  g_{a \gamma \gamma}\dot{\phi} B .
	\label{numerical} 
	\end{align}	
A numerical simulation of the evolution of equations \eqref{numerical}, beginning from a small perturbation in the axion field, is shown in figure \ref{1DFig}.

\pgfdeclareimage[interpolate=true]{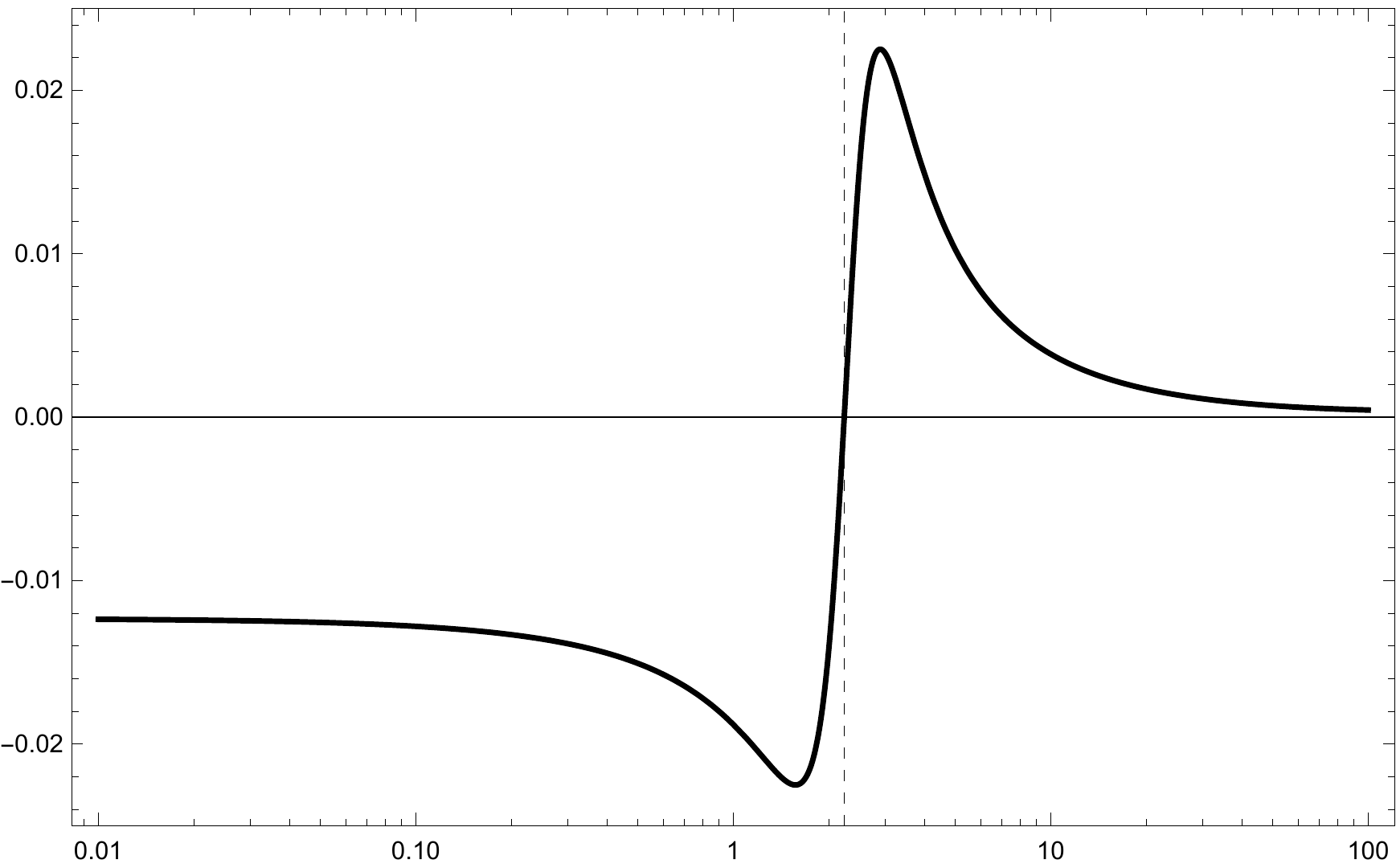}{ImomegaPlot.pdf}
\begin{figure}[ht]
	\begin{center}
		\begin{tikzpicture}[scale=0.4]
		\pgftext[at=\pgfpoint{0cm}{0cm},left,base]{\pgfuseimage{ImomegaPlot.pdf}}
		\node at (10,-0.6) {$k/\mu$};
		\node at (-2,5.7) {$\text{Im}(\omega)/\mu$};
		 \node at (9.4,12) {$c_s =0.01$};
		\end{tikzpicture}
	\end{center}
	
	\caption{The imaginary part of the frequencies for a finite axion sound speed $c_s$ as a function of $k$ for $B g_{a \gamma \gamma}/\mu = 0.3$, $u=0.5$ and $\sigma /\mu = 3$ and $c_s = 0.01$. The dashed vertical line shows $k=k_c$ of eq.~(\ref{KC}). }
	\label{ImOmegaFig}
\end{figure}
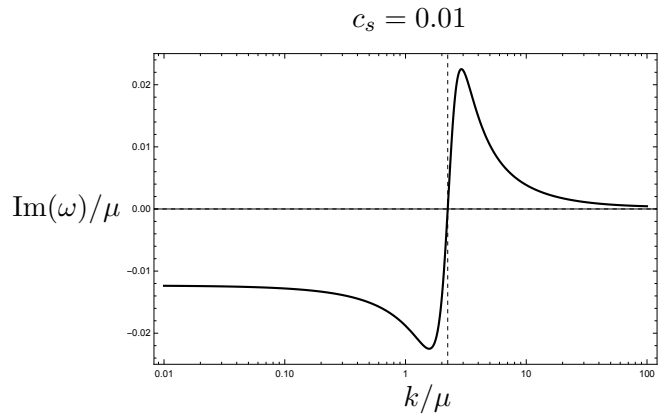
\pgfdeclareimage[interpolate=true]{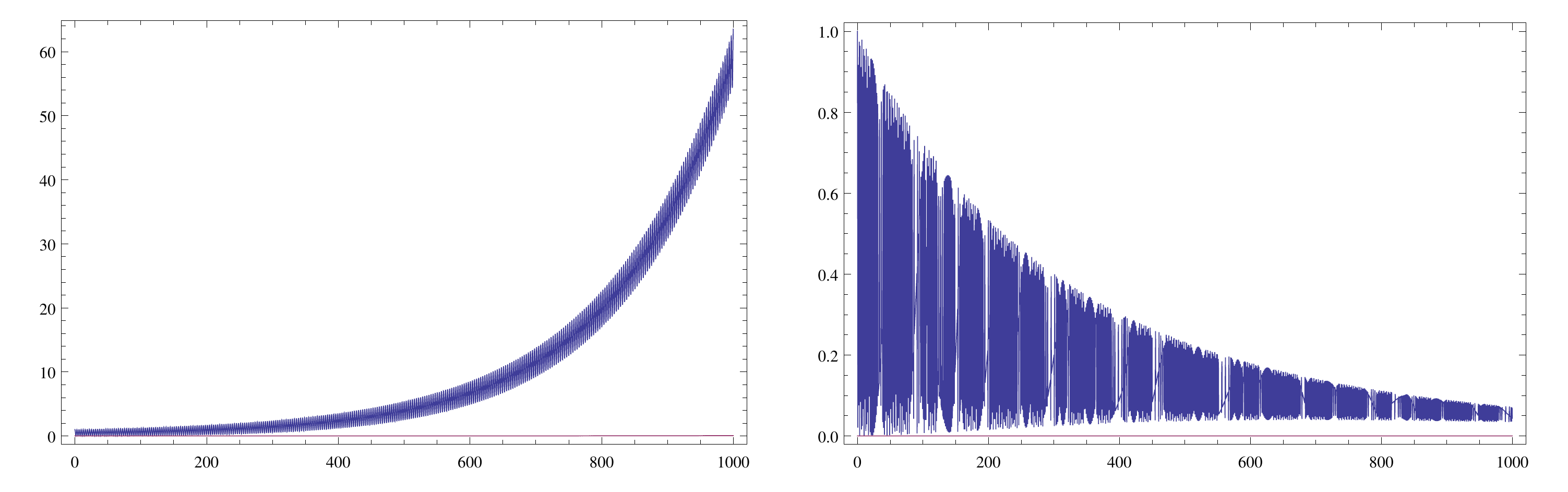}{1DFig.pdf}
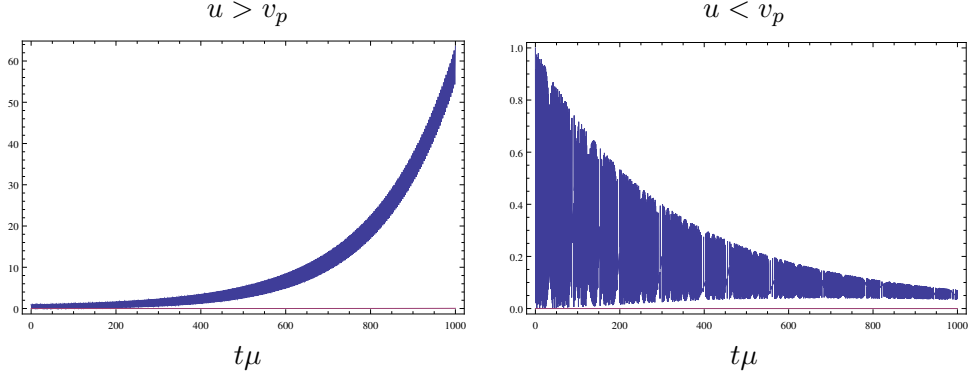
\begin{figure}[ht]
	\begin{center}
		\begin{tikzpicture}[scale=0.44]
		\pgftext[at=\pgfpoint{0cm}{0cm},left,base]{\pgfuseimage{1DFig.pdf}}
		\node at (8,-0.4) {$t \mu$};
		\node at (23,-0.4) {$t \mu$};
		\node at (8,10) {$u > v_p$};
		\node at (23,10) {$u<v_p$};
		\end{tikzpicture}
	\end{center}
	\caption{The axion (blue) and photon component (red) (arbitrary units) for equations \eqref{numerical} for the mode $|\textbf{k}| =3\mu$. The left- and right-hand plots correspond to sound speeds $c_s =0.01$ and $c_s=0.7$, respectively. The other parameter values are $B g_{a \gamma \gamma}/\mu = 0.3$, $u=0.5$ and $\sigma /\mu = 3$. }
		\label{1DFig}
\end{figure}

	\newpage
	\section{Neutron stars and superradiance in 3D}\label{3D}
	To generalise the above discussion to neutron stars, we consider the simplest setup consisting of stationary backgrounds such that the axion and photon fields have a simple harmonic time dependence $\sim e^{-i \omega t}$. This would hold, for instance in the case of an axisymmetric rotating neutron star. The task reduces then to an eigenvalue problem for the discrete frequencies $\omega_{\ell m n}$ associated to the axion boundstates, which satisfy $\omega < \mu$ and have a Yukawa-like confinement $\sim e^{- r \sqrt{\mu^2 - \omega^2}}$ as $r\rightarrow \infty$. Explicitly from eqs.~(\ref{EOM3}) and (\ref{EOM4}), the linearised equations for axion-photon fluctuations in Lorenz gauge about a background plasma with a magnetic field ${\bf B}$ read as follows,
	\begin{align}
	&\square \phi  + \mu^2 \phi = - g_{a \gamma \gamma} \left[ \nabla A^0 + \dot{\textbf{A}}  \right]\cdot \textbf{B},  \nonumber \\
	& \square A^0 =- g_{a\gamma \gamma}\nabla \phi \cdot \textbf{B} - \sigma \textbf{u} \cdot  \left[ \nabla A^0 + \dot{\textbf{A}} \right] ,\nonumber \\
	&\square \textbf{A}  = g_{a \gamma \gamma}\dot{\phi}\textbf{B}  - \sigma \Big[ \nabla A^0 + \dot{\textbf{A}} \Big] + \sigma \textbf{u} \times \left[ \nabla \times \textbf{A} \right]   \label{mixing},
	\end{align}
	where we made the non-relativistic fluid approximation $u^0 = 1$. In general this is a highly non-trivial and multidimensional eigenvalue problem of two mixing fields. In addition, since the spatial dependence of $\textbf{B}(\textbf{x})$, $\textbf{u}(\textbf{x})$ and $\sigma(\textbf{x})$ couple different harmonics of $A^\mu$ and $\phi$, a mode-by-mode treatment for each $(\ell, m, n)$ would seem intractable. Indeed, even in the absence of axion-photon mixing different classes of vector harmonics associated to $A^{\mu}$ are coupled in a non-trivial way for non-vanishing fluid velocity $u^\mu$, as explained in \cite{Cardoso:2017kgn,Pani:2013pma}. 
	 
	 In general it would therefore seem that the only route to the complete eigenspectrum is to solve the full set of coupled partial differential equations (\ref{mixing}) numerically, imposing appropriate boundary conditions for the fields at spatial infinity together with regularity at the origin. 
	 
	Remarkably, however, we find that in certain limits, it is possible to obtain analytic results for the eigenspectrum, which  demonstrate explicitly the existence of superradiant effects. In the following sections we describe the regime for which we have obtained analytic results. Our approach is to treat the axion-photon mixing and the conductivity as a perturbation of a system whose eigenspectrum is already known. The corresponding free eigenfunctions can then be used to construct the eigenstates perturbatively as a double expansion in $g_{ a\gamma \gamma}$ and $\sigma$ in analogy with quantum mechanical perturbation theory \cite{LANDAU1977133,Cohen:2006fh}. 
	
\subsection{A review of axion boundstates}\label{Boundstates}
We review here the basic features of axion-boundstates around gravitating objects, which will prove a useful reference in subsequent calculations. We take the simple Schwarzschild metric
\begin{equation}
ds^2 =  e^{2 \Psi(t)} dt^2 - N^{-1}dr^2 + r^2 \left(d\theta^2 + \sin^2 \theta d \varphi^2  \right),
\label{SchwarzsInt}
\end{equation}
 whose interior form corresponds to a constant density. Explicitly we have
\begin{align}
N(r) & = 
\left \{
\begin{tabular}{cc}
$1- \frac{ r_s r^2}{R^3}$ &  $\qquad \quad r\leq R$\\
$1 - \frac{r_s}{r}$ &   $\qquad  \quad r > R,$
\end{tabular} 
\right. \nonumber \\ 
&\nonumber \\
e^{\Psi} & =  
\left \{
\begin{tabular}{cc}
$\frac{3}{2}\sqrt{1 - \frac{r_s }{R}} - \frac{1}{2}\sqrt{1 - \frac{ r_sr^2}{R^3}}$ &  $\qquad \quad r\leq R$\\
$\sqrt{1 - \frac{r_s}{r}}$ &   $ \qquad  \quad r > R$.
\end{tabular}
\right.  \label{metric}
\end{align}
where $R$ and $r_s = 2 G M$ are the stellar and Schwarzschild radii, respectively. Axion solutions 
\begin{equation}
\phi = \frac{1}{r} \sum_{\ell m} Y_{\ell m} \Phi_{\ell m} (r) ,\label{PhiForm}
\end{equation}
have a radial function $\Phi(r)$ which satisfies a Schr{\"o}dinger-like equation
\begin{equation}
-\frac{d^2 \Phi}{dr_*^2} + 
U_\ell(r)\Phi=  \omega^2 \Phi \label{Schrod},
\end{equation}
with
\begin{equation}
U_{\ell}(r) = 
e^{2\Psi}\left(\frac{l(l+1)}{r^2} + \mu^2\right) + \frac{A' A}{r}
, \qquad \frac{d r_*}{dr} = e^{-\Psi} N^{1/2} \label{U} 
\end{equation}
where $r_*$ is a generalised tortoise coordinate and $A = e^{\Psi} N^{1/2}$. The potential (\ref{U}) will in general admit boundstates, with a quantised set of frequencies $\omega_{\ell n}$, for the cases of interest here, $r_s \mu \ll 1$, outside the star, the potential-well is approximated by
\begin{equation}
U_\ell(r) \simeq \frac{\ell (\ell+1)}{r^2} - \frac{r_s \mu^2}{r} + \mu^2 , \qquad \qquad \mu r_s  \lesssim R\mu \ll 1 \label{Coulomb}.
\end{equation}
One recognises eq.~(\ref{Coulomb}) as precisely the potential associated to the Hydrogen atom with Coulomb interaction $\sim r_s \mu^2/r$. Appropriately normalised bound-states and their associated frequencies are well-known and given by
\begin{align}
&\Phi = \mu^{1/2} \alpha^{1/2}_{\ell n} \sqrt{\frac{n!}{2 (n+ 2 \ell + 1)! (n + \ell +1)}}  e^{-x/2} x^{\ell+1} L_{n}^{2 \ell + 1} 
\left[ x \right], \qquad \omega^2_{\ell n}  = \mu^2 \left( 1 - \frac{\alpha^2_{\ell n}}{4}\right) \label{Hatom},
\end{align}
where
\begin{equation}
x = r \mu \alpha_{\ell n}, \qquad \alpha_{\ell n} = \frac{\mu r_s}{\ell + n  +1}, 
\end{equation}
and $L_n^{2 \ell + 1}(x)$ is a generalised Laguerre polynomial. This solution is in fact a good approximation to the boundstates of eq.~(\ref{Schrod}) in the limit $\alpha \ll 1$ relevant for superradiance around neutron stars. Crucially, the absence of a horizon means that the frequencies are real. The purpose of the remainder of this paper is to ask what happens when one perturbs this boundstate with a magnetic field and rotating magnetosphere.
	
	\subsection{Eigenfrequencies from quantum mechanical perturbation theory} \label{QMPT}
			We shall therefore divide the Hilbert space into a ``free" system consisting of a perfectly conducting spherical star surrounded by vacuum, which is perturbed with an external magnetosphere of conductivity $\sigma_{\! \! _M}$ and axion-photon interaction $g_{a \gamma \gamma}$. The conductivity of the neutron star itself is very large, $\sigma_* \sim 10^{26} \, {\rm s}^{-1}$ \cite{Baiko:1995qg} is typically much larger than any other mass-scale in our setup justifying the assumption of a perfectly conducting star. To proceed with our perturbative treatment we draw an analogy with quantum mechanics and transform the equations of motion from  wave function to operator form via the standard substitutions   $\phi \rightarrow \ket{\phi}$, $A^\mu \rightarrow \ket{A^\mu}$ and $- i \nabla \rightarrow  \hat{p}$ etc.~such that (\ref{mixing}) reads:
\begin{equation}
\left[H(\sigma_*) + V(\sigma_{\! \! _M},g_{a\gamma \gamma}) \right]\left( 
\begin{array}{c}
\ket{\phi}\\
\, \, \ket{A^0} \\
\ket{\textbf{A}}
\end{array}
\right) = \omega^2
\left( 
\begin{array}{c}
\ket{\phi}\\
\, \, \ket{A^0} \\
\ket{\textbf{A}}
\end{array}
\right),
\label{Hamiltonian}
\end{equation}
	with the constraint $\partial_\mu A^\mu =0$ and where $\sigma_{\! \! _M}$ is the conductivity of the magnetosphere and $\sigma_*$ the conductivity of the star, assumed infinite in the stellar interior which gives rise to the condition $\textbf{E} + \textbf{u}\times \textbf{B}=0$ for $r <R$. Note that the fields should not be considered in any sense quantum, and the solutions and eigenvalues derived in what follows correspond to classical field configurations. The bra-ket notation is simply to make the parallel with orbitals encountered in atomic perturbation theory more manifest and provide a compact notation. Neglecting gravitational effects in the photon sector, the free Hamiltonian reads 
	\begin{equation}
	H= 
	\left(
	\begin{array}{ccc}
	- \frac{d^2}{dr_*^2} + U(r) , &\quad 0 & 0 \\
	0 & \quad  - \nabla^2 &  \quad 0\\
	0&\quad 0 &\quad  - \nabla^2 
	\end{array}
	\right), \label{H}
	\end{equation}
	where $U$ is the potential in eq.~(\ref{U}). The perturbation matrix is given by
		\begin{equation}
	V = V_A + V_{a \gamma \gamma} ,
	\end{equation}
	with
	\begin{align}
	V_{a \gamma \gamma}& = i g_{a\gamma \gamma}
	\left( 
	\begin{array}{ccc}
	0  & \quad  \textbf{B}(\hat{x})\cdot  \hat{p} & \quad  - \omega  \textbf{B}(\hat{x})  \\
	\textbf{B} (\hat{x})\cdot \hat{p} & \quad 0 & \quad 0 \\
	\omega \textbf{B}(\hat{x}) & \quad 0 & \quad 0 
	\end{array}
	\right), \label{Vagam}\\
	\nonumber \\
	V_A & =
	i \sigma_{\! \! _M} (\hat{x}) \left(
	\begin{array}{ccc}
	0 &\quad 0 & 0 \\
	0 & \quad \textbf{u}(\hat{x}) \cdot \hat{p} &  \quad  -\omega \textbf{u}(\hat{x}) \\
	0&\quad \hat{p} &\quad  - \omega   - \textbf{u}(\hat{x}) \times \hat{p} 
	\end{array}
	\right) .\label{VA} 
	\end{align}
	We now wish to obtain the eigenvalues of the system perturbatively using the standard method of quantum mechanical perturbation theory \cite{LANDAU1977133,Cohen:2006fh}. Specifically we compute the eigenvalues of the system perturbatively as a double expansion in powers of $g_{a \gamma \gamma}$ and $\sigma_{\!\! _\text{M}}$, starting from the unperturbed frequencies $\omega_{\ell n}$  associated to the boundstate $\ket{\phi_{\ell m n}}$ discussed in sec.~\ref{Boundstates}. We shall take the standard approach of quantum mechanical perturbation theory by expanding the perturbed frequencies $\omega$ as
	\begin{equation}
	\omega = \omega_{\ell n} + \omega^{(1)} + \omega^{(2)} + \omega^{(3)} + \cdots, \label{omegaExpansion}
	\end{equation}
	where superscripts count the perturbation order in $V$, with a similar expansion for the eigenstates. The validity of the above expansion relies on the smallness of the two expansion parameters $g_{a \gamma \gamma}B$ and $\sigma_{_\text{M}}$, in comparison to the only other natural length scale in the problem, $\mu$ (which also sets the frequency scale for boundstates $\omega_{\ell n} \sim \mu$), so that perturbation theory requires $g_{a \gamma \gamma} \left|\textbf{B}\right| \ll \mu$ an $\sigma_{_\text{M}} \ll \mu$.  This is clearly equivalent to the validity of a perturbative expansion of (\ref{dispersion1}) in $g_{a \gamma \gamma}$ and $\sigma$ for length scales set by $k \sim \mu$. One might think that an expansion only in $g_{a \gamma \gamma}$ (which is already constrained to be perturbatively small \cite{Patrignani:2016xqp}) would be more straightforward, with $V_A$ being absorbed into the unperturbed Hamiltonian $H$. However, as alluded to at the beginning of this section, there is no simple set of complete solutions for the photon sector with finite conductivity and fluid velocity \cite{Cardoso:2017kgn}.

	\subsubsection{Unperturbed spectrum of H}\label{H0}
	The first task in quantum mechanical perturbation theory is to find a complete spectrum for the unperturbed Hamiltonian $H$, from which one can construct perturbed eigenfrequencies and eigenstates order-by-order in coupling constants. Let us denote such a basis $\ket{\phi_{\ell m n}}$ and $\ket{A^\mu_{(i)}(\omega),\ell ,m}$, which must satisfy
	\begin{equation}
	H \ket{{A^\mu_{(i)}(\omega),\ell ,m}} = \omega^2 \ket{A^\mu_{(i)}(\omega),\ell ,m}, \qquad \quad H\ket{\phi_{\ell m n}} = \omega^2_{\ell  n }\ket{\phi_{\ell m n}}, \label{Laplace}
	\end{equation}
	where $H$ is given by eq.~(\ref{H}). We have already discussed the discrete spectrum of axion states in sec.~\ref{Boundstates}. The photon states form a continuous spectrum for each quantum number $(\ell, m)$, labelled by a continuous range of frequencies $\omega$ and are therefore non-normalisable. What is important, is that the basis should be complete, such that
	\begin{equation}
	\sum_{\ell \,, m}   \sum_{i} \int d [\omega^2] \ket{A^\mu_{(i)}(\omega, { \bf r}),\ell ,m} \bra{A^\nu_{(i)}(\omega, { \bf r'}),\ell ,m} = \eta^{\mu \nu}\mathbb{I} {\delta^3({\bf r} - {\bf r'})}.  \label{AComplete}
	\end{equation}
	where we use the normalisation of the measure $\int d[\omega^2] =1/2 \int_{-\infty}^\infty d \omega \omega$. Here $i$ corresponds to summing over a complete basis for polarisations, with the remaining sums over $\ell$ and $m$.  There are of course many spectral decompositions of the photon operator (\ref{Laplace}), but here we give one which is most convenient for the calculations which follow in sec.~\ref{Example}. An alternative complete basis constructed from vector spherical harmonics, which would be convenient for general harmonic decompositions of a magnetic field is given in appendix \ref{VecHarmonicBasis}. We therefore construct a complete basis for photon solutions of the form
 \begin{equation}
 A^{(i)\, \ell m }_\mu = \varepsilon^{(i)}_\mu A_\ell (\omega r) Y_{\ell m}(\theta,r),  \label{CartBasis}
 \end{equation}
 where $i = 0,1,2,3$ and $\epsilon_{\mu}^{(i)}$ are constant polarisation vectors with respect to the cartesian coordinates $(t,x,y,z)$ which satisfy
 \begin{equation}
 \sum_{i , \, j} \eta_{i j} \varepsilon^{(i)}_\mu \varepsilon^{(j)}_\nu = \eta_{\mu \nu} . \label{SumA}
 \end{equation}
 In other words, each photon separately satisfies Laplace's equation in Minkowski space. An obvious choice is to take $\epsilon^{(0)} = [1,0,0,0]$, $\varepsilon^{(1)} = [0,1,0,0]$ etc. We also require the radial functions to satisfy
 \begin{equation}
 \int d[\omega^2] A_{\ell}(\omega r) A_\ell(\omega r') = \frac{1}{r^2}\delta(r-r'). \label{completeA}
 \end{equation}
  Here the radial function $A_\ell(\omega r)$ is a solution to
 \begin{equation}
 \left[ \omega^2 + \frac{d^2}{dr^2} - \frac{\ell(\ell+1)}{r^2} \right] (r A_\ell) =0 .\label{Bessel1}
 \end{equation}
 In general the $A_\ell$ will therefore be appropriately normalised linear combinations of  $j_\ell$ and $y_\ell$ - the spherical Bessel functions of the first and second kind, respectively, with the exact form being determined from the boundary conditions $\textbf{E} + \textbf{u} \times \textbf{B}=0$ for $r < R$, corresponding to infinite conductivity in the stellar interior. We show in appendix \ref{axial} that unless the frequencies are an integer multiple of $\Omega$, the $A^{(i)}_\mu$ must vanish in the stellar interior. Therefore, since the $A^{(i)}_\ell$ satisfy Bessel's equation (\ref{Bessel1}) and vanishes inside the star, it must have vanishing Dirichlet boundary condition at $r=R$. The unique solution is given by:
 \begin{equation}
 A^{(i)}_{\ell}(\omega r)  = 
 \frac{1}{N_\ell(\omega R)} \sqrt{\frac{2 \omega}{\pi}} \left[ y_\ell (\omega R) j_\ell (\omega r) -  j_\ell (\omega R)  y_\ell (\omega r)\right] , \quad r \geq R, \qquad \omega \neq m \Omega \label{Ai},
 \end{equation}
 where $N(\omega R)= \left[ j_\ell^2(\omega R) + y_\ell ^2(\omega R)\right]^{1/2}$ gives an appropriate normalisation \cite{10.2307/43638547} to ensure the modes satisfy the completeness relation (\ref{AComplete}) and \eqref{completeA}. This has a discontinuity in $dA_{\ell m}/dr$ corresponding to discontinuities in the associated electromagnetic fields which generate surface currents on the conductor as the modes scatters off the surface. Thus from the orthonormality properties of spherical harmonics, and the relation (\ref{SumA}) it is immediately obvious that the solutions (\ref{CartBasis}) give a complete basis satisfying (\ref{AComplete}).
	
	The purpose of these photon states is to allow one to construct what is in essence a spectral decomposition of the photon Green function. More formally we are interested in the  resolvent 
	\begin{equation}
	G(\omega^2_{\ell n}) \equiv \frac{1}{\mathcal{H} - \omega^2_{\ell n}},
	\end{equation}
	which is used to construct states and eigenvalues perturbatively \cite{Cohen:2006fh}. It is the inverse of the operator $\mathcal{H} - \omega^2_{\ell n}$ such that it satisfies $(\mathcal{H} - \omega^2_{\ell n}) G_{\mu \nu}(\omega^2_{\ell n}) = \eta_{\mu \nu} \delta(\textbf{x},\textbf{x}')$. This can be constructed from a complete set of eigenfunctions
	\begin{equation}
	G_{\mu \nu}(\omega^2_{\ell n}) = \sum_{\ell, m} \sum_i\int d [\omega^2]  \frac{A_{\ell m, \, \mu}^{(i)}(\omega, \textbf{x}) \,A_{\ell m, \, \nu}^{(i)\,*}(\omega, \textbf{x}')}{\omega^2_{\ell n} - \omega^2}, \label{Green}
	\end{equation}
	provided the eigenstates satisfy an appropriate completeness relation.  
	

	\subsubsection{Perturbed eigenfrequencies}
	
	Now that we have discussed a complete spectrum for the unperturbed Hamiltonian, we can use this to construct the eigenvalues perturbatively in $V$. Note that  $V_{a \gamma \gamma}$ is hermitian, since the only non-vanishing component  $(V_{a \gamma \gamma}  \! - \!V_{a \gamma \gamma}^{\dagger})$ is $i g_{a \gamma \gamma}  [\textbf{B}_i(\hat{x}), \hat{p}_i]$. It is easy to see this vanishes since if one takes matrix elements $\bra{\phi} [\textbf{B}_i(\hat{x}), \hat{p}_i] \ket{A^0}$ between any two states, upon integration by parts one can see this is equal to $\int d \textbf{x}^3 \phi^* A^{0} \nabla \cdot \textbf{B}$, which vanishes since  $\nabla \cdot \textbf{B}=0$. Thus we have  $V_{a \gamma \gamma} = V_{a \gamma \gamma}^{\dagger}$. By contrast, $V_A$ is non-hermitian and corresponds to the dissipative plasma dynamics needed for superradiance. We can then construct the eigenfrequencies perturbatively using the standard formulae from quantum mechanical perturbation theory for a perturbation potential $V$ \cite{LANDAU1977133} in accordance with the expansion (\ref{omegaExpansion}). Perturbing about an axion boundstate $\ket{\phi_{\ell m n}}$, we find a contribution to to $\text{Im}(\omega)$ from the finite conductivity at third order in perturbation theory given by 
	\begin{align}
	&\delta \omega_{\ell m n}  = \nonumber \\
	&\frac{1}{2 \omega_{\ell n}}  \sum_{\ell_{1,2},m_{1,2}} \sum_{i,j}\int d [\omega^2_{1,2}]
	\frac{ \bra{   \phi_{\ell m n}  } V \ket{  A^{(i)}_{\ell_1 m_1}(\omega_1) } 
		\bra{A^{(i)}_{\ell_1 m_1}(\omega_1)  }V \ket{ A^{(j)}_{\ell_2 m_2 }(\omega_2)} 
		\bra{A^{(j)}_{\ell_2 m_2 } (\omega_2)  } V  \ket{\phi_{\ell m n}} }
	{ \left(\omega_{\ell n}^2 -\omega^2_1 \right) \left( \omega_{\ell n}^2 -\omega^2_2 \right)    }. \label{deltaOmega}
	\end{align}
	 Note the $1/2\omega_{\ell n}$ prefactor comes from extracting the first order perturbation from $\omega^2 = (\omega_{\ell  n} + \omega^{(1)} + \cdots)^2$, which is precisely $2 \omega_{\ell n} \omega^{(1)}$. Taking residues gives
	\begin{align}
	&\delta \omega_{\ell m n} = \frac{ \pi^2}{8 \omega_{\ell n}}  \sum_{\ell_{1,2},m_{1,2}} \sum_{i,j} \nonumber \\
	&\bra{   \phi_{\ell m n}  }  V_{a \gamma \gamma} \ket{  A^{(i)}_{\ell_1 m_1}(\omega_{\ell n}) } 
	\bra{A^{(i)}_{\ell_1 m_1}(\omega_{\ell n})  }  V_A \ket{ A^{(j)}_{\ell_2 m_2 }(\omega_{\ell n})} 
	\bra{A^{(j)}_{\ell_2 m_2 } (\omega_{\ell n})  } V_{a \gamma \gamma} \ket{\phi_{\ell m n}}  \label{omega3},
	\end{align}
	where we have made explicit which components (\ref{Vagam}) and (\ref{VA}) contribute to each matrix element. Note that in (\ref{omega3}), to leading order in perturbation theory, $V= V(\omega_{\ell n})$ is evaluated at the unperturbed frequency $\omega =\omega_{\ell n}$.
	
	 Within the context of perturbation theory, the problem of determining whether or not a particular configuration $\textbf{B}$, $\textbf{u}$ and $\sigma$ exhibits superradiance, reduces to computing the matrix transition elements in (\ref{omega3}). For a given $V$, only transitions between axion and photon states separated by specific quantum numbers are permitted, meaning that only certain terms in the sum (\ref{omega3}) will contribute, in analogy to selection rules for atomic transitions.

	 \subsection{Example: constant magnetic field with a co-rotating fluid}\label{Example}

	 Evaluating the matrix transition elements for a magnetic field with a general vector harmonic structure would be rather involved. However, for a constant magnetic field, azimuthal fluid velocity and spherically symmetric magnetosphere conductivity
	 \begin{equation}
	 \textbf{B} =B \hat{\textbf{z}} , \qquad u^\mu =(1,0,0,\Omega), \qquad \sigma_{\! \! _M}=\sigma_{\! \! _M}(r) \label{conditions},
	 \end{equation}
	 the result simplifies greatly. Note $\hat{\textbf{z}}$ is the unit vector in the z-direction.  We now compute the matrix elements associated to this setup. We give an example calculation of the matrix elements for $A^{(3)}$ in appendix \ref{EigenCalculation} with the remaining matrix elements following along similar lines.   In the case \eqref{conditions}, the axion mixes only with the $A^{(0)}$ and $A^{(3)}$ basis components, explicitly the only non-vanishing mixing elements for $V_{a \gamma \gamma}$ to leading order in $R \omega$ are:
	  \begin{align}
	   \bra{   \phi_{\ell m n}  }  V_{a \gamma \gamma} \ket{  A^{(3)}_{\ell' m'}} &=- i  \delta_{m m'}\delta_{\ell \ell'}  g_{a\gamma \gamma} B\mathcal{N}_\phi
	   \left(\frac{\omega}{\alpha \mu} \right)^{3/2}\,I[\ell, \ell+1],
	  \end{align}
	  and
	\begin{align}
	   &\bra{   \phi_{\ell m n}  }  V_{a \gamma \gamma} \ket{  A^{(0)}_{\ell' m'}}  \nonumber \\
	   &= g_{a\gamma \gamma} B \mathcal{N}_\phi\left(\frac{\omega}{\alpha \mu} \right)^{\! \! 3/2} \! \!\!\! \delta_{m m'}\Bigg[ S[\ell,\ell',m] I[\ell, \ell'] + \frac{N_{\ell - 1, m}}{N_{\ell,m}} \left(\frac{\alpha \mu}{\omega} \right)( |m| \! - \! \ell) \delta_{\ell[\ell'+1]} I[\ell\!-\!1 , \ell]  \Bigg] . \label{Va0}
	  \end{align}
	  where $\mathcal{N}_{\phi} =  \sqrt{n!}[\pi(n+ 2 \ell + 1)! (n + \ell +1)]^{-1/2}$ is the axion normalisation factor and we have suppressed the subscripts on $\alpha_{\ell n}$ and $\omega_{\ell n}$. $N_{\ell, m} = \sqrt{\frac{(2 \ell+1)(\ell-|m|) !}{4 \pi (\ell+|m|) !}}$ is the normalisation of the spherical harmonics. The integral $I[\ell_1,\ell_2]$ arises from the inner product of the axion and photon radial functions. To leading order in $\omega R$ the solutions (\ref{Ai}) are given by $A^{(i)}_{\ell} \simeq \sqrt{2 \omega/\pi} j_{\ell}(\omega r)$, and using the approximation \eqref{Hatom} for the axion boundstates we have
	  \begin{equation}
	     \text{I}[\ell_1, \ell_2] =  \int_0^\infty dx \, \Bigg[e^{-x/2}  x^{\ell_1+2} L_{n}^{2 \ell_1 + 1}  
	     \left(   x \right)\Bigg]j_{\ell_2-1} \left(\frac{\omega}{\mu} \frac{x}{\alpha} \right) . \label{Ill}
	  \end{equation}
	  Noting that $\mu /\omega \simeq 1$, this can then be expanded as  a power series in $\alpha$. For instance, $I[\ell, \ell +1] \simeq \alpha^{\ell+4} 2^\ell (\Gamma[2 + \ell] \Gamma[2 + 2 \ell + n])/(\Gamma[2 + 2 \ell] \Gamma[\ell + n])$.	  Notice that the various Kronecker deltas and the function
	   \begin{equation}
	   S[\ell, \ell', m] = 
	   \sqrt{(2 \ell +1 )( 2 \ell' + 1)}
	   \left( 
	   \begin{array}{ccc}
	   \ell & \ell' & 1\\
	   0  &   0 &    0
	   \end{array}
	   \right)
	   \left( 
	   \begin{array}{ccc}
	   \ell & \ell' & 1\\
	   -m  &  m  &    0
	   \end{array}
	   \right),
	   \end{equation}
	   where (...) are the Wigner 3-j symbols, describe the selection rules for the magnetic field (\ref{conditions}) and controls the strength of coupling between different spins. Explicitly, $S[\ell, \ell', m]$ is non-vanishing for $\ell - 1 \leq \ell' \leq \ell + 1$, whilst the $\delta_{\ell[\ell'+1]}$ factor allows transitions $\ell \rightarrow \ell-1$. In other words, $\textbf{B}=B\textbf{z}$ allows transitions between spins $\ell \rightarrow \ell \pm 1$. Note however that $m$ is conserved in all cases since $\textbf{B} = B \hat{\textbf{z}}$ is axisymmetric.	  
	  
	  Next one must compute the plasma matrix elements associated to $V_A$ which describe the superradiant scattering of photons. The convergence of the integrals associated to $V_A$ depends on the decay of $\sigma_{\! \! _M}$.  We make a simple model and assume constant conductivity within the light-cylinder, and vanishing conductivity outside so that $\sigma_{\! \! _M} = \text{constant}$ for $R \leq r \leq R_{\text{LC}} = \Omega^{-1}$ and zero otherwise, where $R_\text{LC}$ is the radius of the light-cylinder. To leading order in $\omega R$ and $\omega R_{_{LC}}$ the three relevant matrix elements are
	     \begin{align}
	     \bra{   A^{(0)}_{\ell m}  }  V_A  \ket{  A^{(0)}_{\ell' m'}}&  \simeq    i  \sigma_{_\text{M}}   m \Omega  \frac{  (R_{ _\text{LC}} \omega)^{2\ell+3}- (R \omega)^{2\ell+3}}{ \omega^2  2^{2(\ell+1)} }\frac{1}{\left(l+\frac{3}{2}\right) \Gamma \left(l+\frac{3}{2}\right)^2} \, \delta_{m m'} \delta_{\ell \ell'} \label{V00} ,\\
	      \nonumber \\
	     \bra{A^{(3)}_{\ell m } } V_A \ket{ A^{(3)}_{ \ell' m' }} &\simeq i \sigma_{_\text{M}}  \left( m \Omega - \omega  \right) \frac{  (R_{ _\text{LC}} \omega)^{2\ell+3}- (R \omega)^{2\ell+3}}{ \omega^2  2^{2(\ell+1)} }\frac{1}{\left(l+\frac{3}{2}\right) \Gamma \left(l+\frac{3}{2}\right)^2}  \delta_{m m'} \delta_{\ell \ell'}, \label{V33} 
	     \end{align}
	     with the third corresponding to mixing between the two photon modes:
	     \begin{align}
	     &\bra{   A^{(3)}_{\ell m}  }  V_A  \ket{  A^{(0)}_{\ell' m'}} \simeq  \frac{\sigma_{_\text{M}} \delta_{m m'}}{ \omega^2 2^{ \ell}\Gamma[\ell +3/2]} \nonumber \\
	     &\times \left[  \frac{S[\ell,\ell',m]  [(\omega R_{_{LC}})^{\ell + \ell' +2}\! \! -(\omega R)^{\ell + \ell' +2}]}{2^{\ell'}(2 + \ell + \ell')\Gamma[\ell' +1/2]}
	     + \frac{N_{\ell - 1, m}}{N_{\ell,m}} \delta_{\ell [\ell '+1]}\frac{(|m| - \ell) [(\omega R_{_{LC}})^{2 \ell +1} \! \! -(\omega R)^{2 \ell +1}]}{2^{\ell+1}\Gamma[\ell + 3/2]}
	     \right]. \label{V30}
	     \end{align}
	     Note that since $\textbf{u}$ is perpendicular to $\textbf{A}^{(3)}$, which is polarised in the z-direction, the matrix element $
	     \bra{   \phi_{\ell m n}  }  \omega \,\textbf{u} \cdot \ket{  \textbf{A}^{(3)}_{\ell' m'}}$ vanishes and therefore does not contribute.
	     
	     We see therefore that the factor $\sigma_{_M} (m \Omega - \omega)$ appearing in (\ref{V33}) gives rise to superradiance. Consider the special case $\ell = m$, for which the second terms vanish in (\ref{V30}) and (\ref{Va0}). Furthermore, the matrix element \eqref{V00} appears at $\ell' = \ell+1$ and is therefore angular momentum suppressed by an additional factor $(\omega R_{LC})^2$. The contribution from substituting the remaining matrix elements into \eqref{omega3} gives, for the $\ell=m$ case
	      \begin{equation}
	      \text{Im}\left[\omega_{\ell m n} \right]\simeq  \pi
	       g_{a \gamma \gamma}^2 B^2\sigma_{_\text{M}}  \Bigg( \!\!\left( m \Omega - \omega  \right) \! - \! \omega \, S[\ell, \ell+1, m ]^2\Bigg)  \frac{  (R_{ _\text{LC}} \omega)^{2\ell+3}- (R \omega)^{2\ell+3}}{32 \omega^3   } \alpha^{2 \ell +5} \mathcal{F}_{\ell n}  \label{EvalFinal1},
	      \end{equation}
	      where we expanded (\ref{Ill}) to leading order in $\alpha$ and where
	      \begin{equation}
	      \mathcal{F}_{\ell n}=\frac{2 \Gamma (l+2)^2 \Gamma (2 l+n+2)}{(2 l+3) n! (l+n+1) \Gamma \left(l+\frac{3}{2}\right)^2 \Gamma (2 l+2)^2}.
	      \end{equation}
	      The factor $S[\ell,\ell+1 , m=\ell]^2 = 1/(3 + 2 l)$ arises from the transitions $\ell=1 \leftrightarrow \ell=2$ between $A^{(0)}$ and $A^{(3)}$ and represents a suppression of the coupling between different spins. Thus although the factor $-\sigma \omega S[\ell, \ell+1, m]$ in  \eqref{EvalFinal1} reduces the superradiance rate, it is spin-suppressed.

	   The characteristic time scales associated to the instabilities $\tau_{_I} = 1/\text{Im}(\omega)$ for the case \eqref{EvalFinal1} is plotted in fig.~\ref{FreqPlot} and compared to the characteristic age of the pulsar J1748-2446ad \cite{Hessels:2006ze} given by its spin down time 
	   \begin{equation}
	   \tau_\text{age} = \Omega/\dot{\Omega}.
	   \end{equation}
	   Note this describes the rate of loss of rotational energy $E_{\rm rot} = I\Omega^2/2 $ where $I$, assumed constant, is the moment of inertia of the neutron star. It then follows that
	   \begin{equation}
	   \frac{d E_\text{rot}}{d t} =  - \frac{2}{\tau_{\text{age}}} E_\text{rot},
	   \end{equation}
	   so that $\tau_{\text{age}}$ indexes the rate of angular momentum loss. 
	   
	    The spin down time can be very long: $10^5-10^9 \text{yr}$ for radio pulsars and can be as high as $10^9-10^{11}\text{yr}$ for X-ray pulsars \cite{Harding:2013ij}. Note that the curve in fig.~\ref{FreqPlot} corresponding to constant $\textbf{B}$ uses the value at the stellar surface $B=B_*$. In reality $B$ will fall-off as $\sim (R/r)^3$, this modifies the radial integrals for the axion-photon mixing in such a way that the leading order behaviour, in e.g the $\ell=m=1$ changes from $\sim \alpha^7$ to $\sim (\mu R)^6 \alpha^4$, which is typically a few orders of magnitude lower. 
	    
We have calculated the superradiant timescale for the simple case of a constant magnetic field in the $z$ direction. We might wonder how our results would differ if we had used a more realistic dipole field configuration. For an aligned rotator configuration, in which the pulsar's magnetic and rotational axes are aligned, the magnetic field is:

\begin{equation}
\label{dipole}
{\bf B} ({\bf r}) = \frac{B_0}{r^3} \left( 2 {\rm cos} \theta \, \hat{\bf r} + {\rm sin} \theta \, \hat{\boldsymbol \theta} \right) = \frac{B_0}{(\rho^2 + z^2)^{3/2}} \left( \frac{3 z \rho}{z^2 + \rho^2} \hat{\boldsymbol \rho} + \frac{2 z^2 - \rho^2}{z^2 + \rho^2} \hat{\bf z} \right),
\end{equation}

\noindent where $\bf B$ is expressed in spherical polar and cylindrical coordinates. The axion eigenstates that are active for superradiance are primarily concentrated in a torus around the equator of the neutron star, as defined by the rotational axis. For the $l=|m| = 1$ orbitals, we have:

\begin{equation}
Y^1_1 \sim {\rm sin} \theta.
\end{equation}	     		
	
\noindent Near the equator ($\theta \sim \frac{\pi}{2}$, $z \ll \rho$) where the axion is concentrated, the magnetic field in equation \eqref{dipole} is approximately in the $z$ direction. We therefore expect a constant field in the $z$ direction to be a good approximation for pulsars where the magnetic and rotational axes are approximately aligned, and that the fractional corrections induced by the curvature of the dipole field will be $\mathcal{O} (1)$.
	     
	 \pgfdeclareimage[interpolate=true]{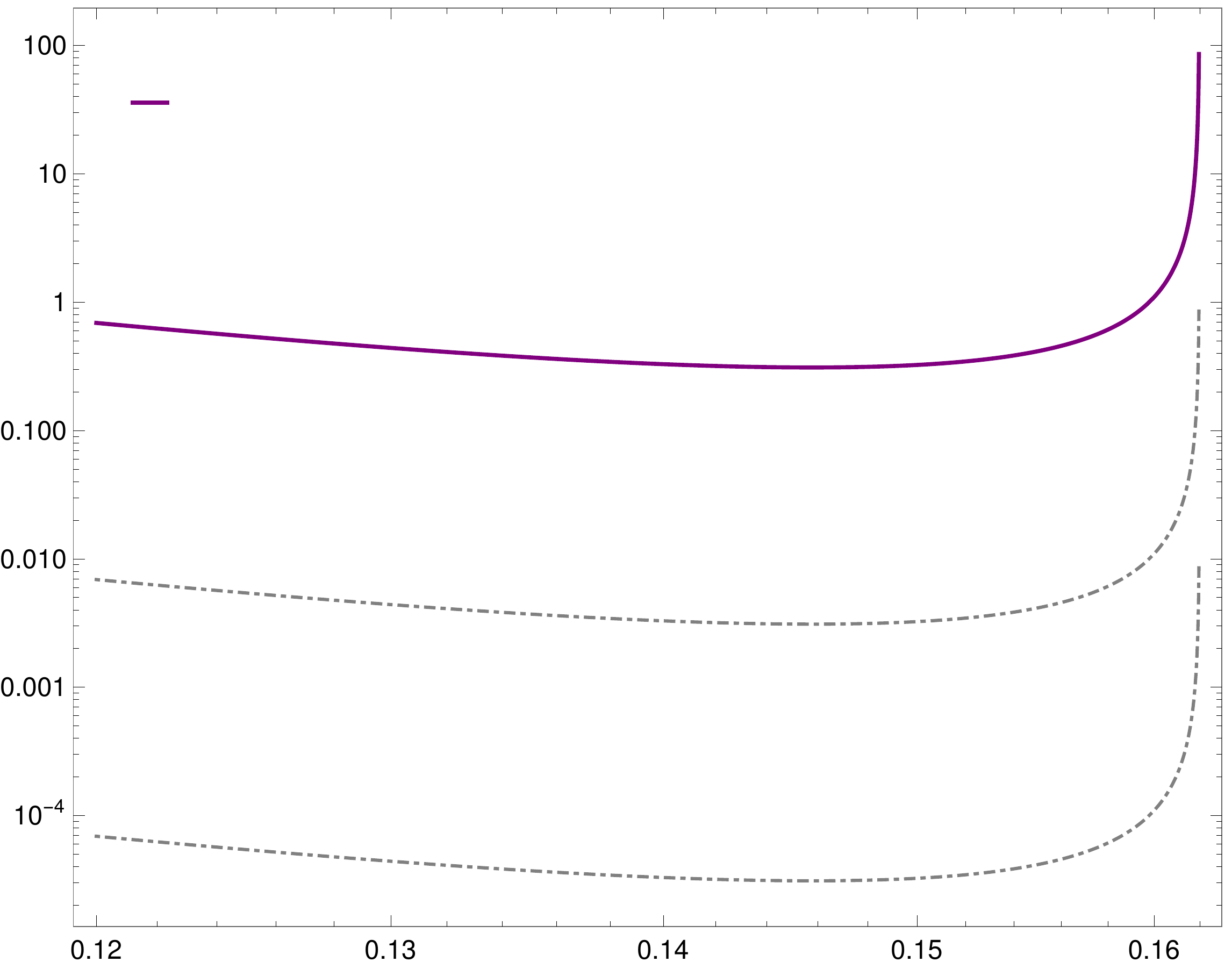}{FreqPlots.pdf}
	 \begin{figure}[ht]
	 	\begin{center}
	 		\begin{tikzpicture}[scale=0.4]
	 		\pgftext[at=\pgfpoint{0cm}{0cm},left,base]{\pgfuseimage{FreqPlots.pdf}}
	 		\node  at (5.7,16.4) {\footnotesize  J1748-2446ad };
			\node at (12.5,10.5) {\tiny $B=10^{9} \text{G}$};	 		
	 		\node at (12.5,6) {\tiny $B=10^{10} \text{G}$};
	 		\node at (12.5,2.4) {\tiny $B=10^{11} \text{G}$};
	 		\node at (12.5,-0.7) {\small  $\mu \, R$};
	 		\node[rotate=90] at (-1.5,8.5) {\small $\tau_{_I}/\tau_{\text{age}} \,\, $};
	 		\end{tikzpicture}    	
	 		\caption{Superradiant instability time-scales  $\tau_{_I} = 1/|\text{Im}(\omega)|$  from eq.~\eqref{EvalFinal1} with $\ell=m=1$ relative to the spindown time scale $\tau_\text{age}=\Omega/\dot{\Omega}$ of the star. We neglected the spin-spin interaction term. We used the pulsar J1748-2446ad  and took $g_{a \gamma \gamma} = 10^{-12} \text{GeV}^{-1}$, with a conductivity for the magnetosphere $\sigma_{\! \! _M} R=0.1$. For the neutron star we use the values $R=10\text{km}$ and $M=1.4 M_\odot$. We also show the same rate with different magnetic field values in order to illustrate the dependence of the mixing strength.}
	 		\label{FreqPlot}
	 	\end{center}
	 \end{figure}
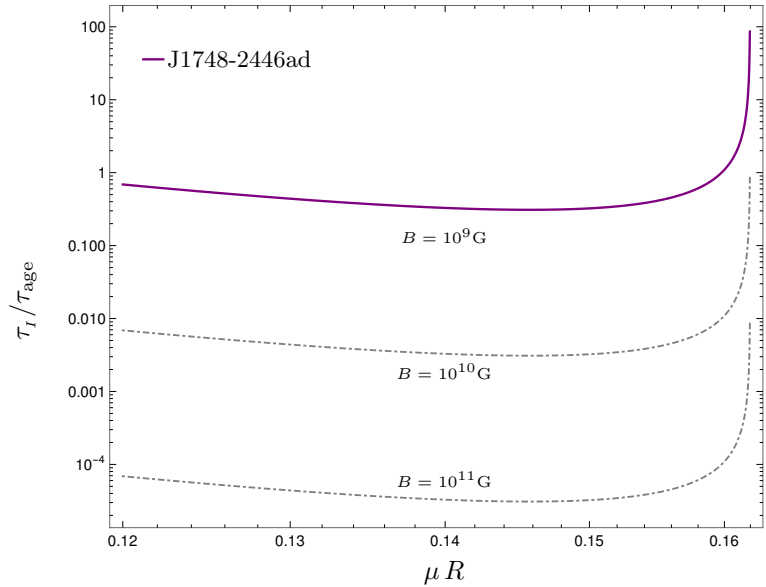

	\section{Nature of conductivity in neutron star magnetospheres}\label{Conductvity}
	
	The results of this paper are intended to illustrate the potential for instabilities to develop when ultra-light fields interact with low-frequency plasma modes in the spirit of \cite{1811.04945,Ikeda:2019fvj} and should be considered only as the first steps towards a more comprehensive analysis. In this sense, we hope to motivate a study into wider classes of axion-plasma instabilities which might draw on the rich class of unstable modes in neutron stars, e.g.~\cite{Andersson:2000mf}. See also interesting developments in \cite{Tercas:2018gxv,Mendonca:2019eke}.
	
	In particular, our perturbative treatment of the conductivity, valid when $\sigma \lesssim \omega$ is intended only to provide the first analytic insights into such effects and is not an exhaustive study across the full range of conductivities in neutron stars or their magnetospheres, which will be the subject of future work \cite{McDonald}.
	
	Nonetheless, it is interesting to expand a little more on the nature of conductivities in neutron star magnetospheres relevant for global magnetospheric perturbations which lead to the axion superradiance presented here. There are two main points to make in this respect. The first is that unlike the conductivity associated to the dark-photon superradiance of ref.~\cite{Cardoso:2017kgn}, which is essentially a free parameter, the conductivity here relates to the global background structure of the magnetosphere and not that of an associated dark sector. As a result, the value of $\sigma$ is constrained by matching pulsar magnetosphere models with finite conductivity \cite{Li:2011zh, Kalapotharakos:2011vg} to observation \cite{Kalapotharakos:2013sma,Brambilla:2015vta} leading to typical values $0.01 \Omega \lesssim \sigma \lesssim 100 \Omega$. Secondly, it is important to realise that the global nature of conductivity in neutron star magnetospheres is quite  different in character to that encountered for local fluctuations in plasmas.

	In studying plasmas in other astrophysical contexts  the usual procedure \cite{Cardoso:2017kgn, Conlon:2017hhi,Sen:2018cjt,1811.04945} is to assume a Drude model, $\sigma_{\rm Drude} (\omega) = \frac{n_e  e^2 \tau_\text{coll}}{m_e (1 - i \tau_\text{coll} \omega)}$, where $\tau_{\rm coll}$ is the average time between collisions of charge carriers, $n_e$ is the number density of electrons, and $m_e$ the electron mass. The dissipation of electric fields in the medium is then described by ${\rm Re}(\sigma_{\rm Drude})$. However, whilst this conductivity is appropriate for local fluctuations (i.e.~WKB approximation) of the magnetosphere, here we deal with global multipole perturbations of the magnetosphere itself, which result simply from linearising global perturbations of the magnetosphere about the background equations analgous to those in\footnote{We use a slightly different form of Ohm's law here, but the logic remains the same.} \cite{Li:2011zh, Kalapotharakos:2011vg}. Therefore by definition, one must use the conductivity appropriate to the global structure of magnetosphere solutions. 
	
	Although these models are somewhat phenomenological, one can nonetheless correlate the values of the conductivity $\sigma$ to the properties of pulsars both via observation and on the basis of a broad theoretical picture we describe below. This ``global" conductivity, far from being arbitrary, is intimately related to the macroscopic accelerating background fields. Specifically it models the inability of plasma to completely screen $\textbf{E}\cdot \textbf{B}$ \cite{Li:2011zh} and parametrises the departure from the ideal magnetohydrodynamics or ``force-free" condition $\textbf{E} + \textbf{v} \times \textbf{B} =0$. These accelerating fields, characterised by so called \textit{acceleration gaps} in which $\textbf{E}\cdot \textbf{B}$ is non-vanishing are vital for reproducing the observed emission spectra of pulsars since the resulting acceleration of charges along magnetic field lines produces observed emission spectra. If $\sigma$ is everywhere too high, such acceleration cannot happen. By matching to observation, one can determine that larger values of $\sigma$ correspond to higher spin-down luminosity and therefore shorter ages of the pulsars as shown in figs.~3 and 4 of ref.~\cite{Brambilla:2015vta}. Thus faster millisecond pulsars, which have the longest spin down, times motivate lower global conductivities, and also produce the strongest superradiance effects owing to their high spins. Indeed, even if there are smaller ``dissipation regions" \cite{Kalapotharakos:2017tzm} where $\sigma$ takes lower values, the axion tail $\phi \sim r^{\ell}$ can still interact with unsuppressed electromagnetic modes in these regions giving a non-zero axion-photon overlap integral, leading to superradiance.  
	

	One can also gain important physical insight into the allowed values of $\sigma$ through simple dynamical arguments. From Ohm’s law we have $\textbf{J} = \sigma (\textbf{E} + \textbf{u} \times \textbf{B})$ and since for neutron stars, $\textbf{B}$ is typically much larger then $\textbf{E}$, the Lorentz force per unit volume is therefore $\textbf{J} \times \textbf{B}$ and of order $\sigma u B^2$.  By contrast, the Coriolis force is given by $2 \rho \boldsymbol{\Omega} \times \textbf{u}$. The ratio of these two forces, is characterised by the \textit{Elsasser number} \cite{PhysRev.70.202,Li:2011zh,2007egp..book.....G}, $\Lambda$:
	\begin{equation}
	\Lambda =\frac{\sigma B^2}{\rho\Omega },
	\end{equation}
	where $\rho$ is the mass-density of the plasma. From this we see that $\sigma$ is important in determining the size of $\Lambda$. 	One can also relate $\Lambda$ to the the Alfv{\'e}n velocity $v_A =B/\sqrt{\rho}$
	of the plasma, from which one can write
	\begin{equation}
	\Lambda = \frac{v_A^2 \sigma}{  \Omega} .
	\end{equation}
	 By matching pulsar magnetosphere models with finite $\sigma$ to observation,\cite{Kalapotharakos:2013sma,Brambilla:2015vta}, one typically finds $0.01 \Omega \lesssim \sigma \lesssim 100 \Omega$, indicating that the Lorenz and Coriolis forces are in approximate balance with $\Lambda$ not more than a few orders of magnitude from unity. We thus see that the value of $\sigma$ interpolates between two extremes \cite{Melrose:2016kaf} of pulsar magnetosphere models. Empirically it is found that $\sigma \ll \Omega$ corresponds to the so-called vacuum dipole model \cite{Deutsch} consisting of a rotating magnetic dipole surrounded by empty space. Conversely, in the large-$\sigma$ regime one obtains the so-called ``force free" limit of the magnetosphere. Realistic pulsars lie somewhere between these two extremes, explaining why $\Lambda$ is empirically observed to be not too far from unity. 
	 
	  
	  One could also ask to what extent an imaginary part of the magnetospheric conductivity $\text{Im}(\sigma_{\! \! M})\neq 0$ would affect our results. Indeed one might worry about the photon developing a plasma mass via $\text{Im}(\sigma_{\! \! M})\neq 0$ which could prevent superradiant photon scattering unless the threshold condition $\omega_{\text{pl}} < \Omega$ is satisfied. However this is not the case, as we now explain. Clearly the formula (\ref{EvalFinal1}) for the imaginary part of the frequency is unaffected by the imaginary part of the conductivity and since this formula is consistent for sufficiently small $\sigma$, provided the perturbativity condition $\text{Im}(\sigma_{\! \! M}) < \mu$ is satisfied, to leading order in perturbation, the linear $\sigma$  result (\ref{EvalFinal1}) is unaffected by the imaginary part of the conductivity. Similarly from the formula (\ref{dispersion1}) we see that (setting $\textbf{E}=0$), to linear order in $g_{a \gamma \gamma}$
	  \begin{equation}
	  \text{Im}(\omega)  = - \frac{\sigma_R (k\cdot u) g^2_{a \gamma \gamma} }{2(\textbf{k}^2 + \mu^2)^{1/2}}\left[ \frac{ \omega^2 |\textbf{B}|^2-(\textbf{k} \cdot \textbf{B})^2 }{\mu^4 +  (\sigma_I^2 + \sigma_R^2) (k \cdot u) }\right],
	  \end{equation}
	  where $\sigma_R$ and $\sigma_I$ are the real and imaginary parts of the conductivity, respectively. Note that the imaginary part has no-bearing on the overall sign of $\text{Im}(\omega)$ and gives no threshold condition on its overall sign. Instead, $\sigma_I$, only provides a suppression of the overall size of the result. Thus supposing for some reason that one should include an imaginary part of the conductivity in modelling the magnetosphere as in refs. \cite{Li:2011zh, Kalapotharakos:2011vg,Kalapotharakos:2013sma,Brambilla:2015vta}, then provided $\sigma_I$ does not greatly exceed $\sigma_R$, it should have no significant quantitative bearing on the superradiance rate presented here. 
	  
	  
	   One can also study the effect of an imaginary component of $\sigma_{\! \! M}$ by examining the superradiant scattering of the axial mode (see appendix \ref{VecHarmonicBasis}) in a conducting medium. As shown in ref. \cite{Cardoso:2017kgn} to leading order in spin-spin coupling between different photon modes, it satisfies the equation:
	  \begin{equation}
	  \left[- \frac{d^2 }{dr^2} + \frac{\ell(\ell+1)}{r^2} - i \sigma (\omega - m \Omega)  \right](r A_{\ell m}) = \omega^2 (r A_{\ell m}).
	  \end{equation}
	  Clearly in the superradiant regime $\omega < m \Omega$, an imaginary part of the conductivity gives rise to a tachynoic mass-squared $\mu_{\text{tac}}^2 = - \sigma_I (m \Omega - \omega) $, which does not affect superradiant scattering, which is only threatened by \textit{real} masses satisfying $\mu \gtrsim \Omega$. This shows again how the imaginary part of $\sigma$ does not affect the threshold condition for superradiance in the present setup. Instead we would expect it to provide only suppressing factors, as can be extrapolated from eq.~(19) of \cite{Cardoso:2017kgn}.

	  This peculiarity in the present setup is due to the fact that both the photon mass term and the source of rotational energy are derived from the plasma. By contrast, in the case of a black hole, the rotation comes from the black hole horizon, which has velocity $\Omega_H$ which can in general be independent of the photon mass sourced by an external plasma. In that case, one would need to impose $\omega_{\text{pl}} < \Omega_H$ as an additional constraint for superradiance to occur, as considered in \cite{Conlon:2017hhi}.

	  \subsection{Larger conductivities}
	  It is interesting to speculate on what happens in the large conductivity limit characterised by $\sigma \gg \omega, \mu$ as would happen in the stellar interior, or magnetosphere models with larger values of $\sigma$. For large conductivities, the rotating plasma becomes a good conductor such that in its rest frame there can be no-electric field. Thus only thus electromagnetic fluctuations which approximately co-rotate with the plasma are permitted as alluded to in appendix \ref{axial}, leading to a narrow resonance around $\mu, \omega \sim \Omega$, which, as found in fig.~1 of ref.~\cite{Cardoso:2017kgn} has a superradiance rate whose peak value is independent of $\sigma$, with the width narrowing for increasing $\sigma$. This corresponds to the fact that the superradiant term always appears in the combination $\sigma (m \Omega - \omega)$ and thus large values of $\sigma$ are compensated by tuning $\omega \simeq m \Omega$ giving rise to a resonant-like effect. This scenario will be investigated in a subsequent publication \cite{McDonald} where the full range of conductivities will be explored.  
	  
%

	\section{Discussion}\label{Discussion}
	In this paper, we have described a new form of instability occurring outside neutron stars, in which mixing macroscopic axion and photon modes extract rotational energy from the neutron star magnetosphere via a finite bulk conductivity. We derived an explicit expression for the superradiant frequencies in the small conductivity limit, using quantum mechanical perturbation theory, which demonstrates explicitly their existence and the nature of the mechanism.  
	
	In principle, the next steps would be to attempt to derive analytic results also in the large conductivity limit. One should also consider more general configurations for the background magnetic fields than those considered in sec.~\ref{Example}, for instance a dipolar configuration. This would entail more complicated selection rules and mixing of different quantum numbers than for a constant vertical magnetic field. After that one should derive the eigenfrequencies by solving the equations (\ref{mixing}) numerically with appropriate boundary conditions for the fields. This would allow for a comparison of the analytic and numerical results.  
	
		Although we deal with a specific form of non-hermitian dynamics arising from the real part of the conductivity in an Ohm-type law, the result here is clearly an example of a more general phenomenon which can arise when there is an instability in the plasma sector, as is apparent from eq.~(\ref{omega3}). Specifically, for axion modes which couple to an unstable mode of the neutron star -- of which there are many \cite{Andersson:2000mf}-- one could in principle find similar instabilities. Suppose that the axion interacts with a mode $(\ell,m)$ in the plasma (not necessarily associated to $A_\mu$, but to e.g. velocity or density fluctuations) whose effective Hamiltonian $H_A$ contains non-hermitian part $i \Gamma$ such that $i\bra{\ell,m}\Gamma \ket{\ell , m}$ gives an imaginary contribution. Then already at the perturbative level one could imagine a relation similar to (\ref{omega3}) which would give rise to an axion-plasma instability.  It would therefore be of interest to see if there is a wider class of instabilities arising from the mixing of the axion with other unstable modes in the plasma of the neutron star or its magnetosphere.

		\acknowledgments
		JIM is grateful to the support of a Technical University Foundation Fellowship and latterly an Alexander von Humboldt Fellowship. He would also like to thank Carlos Tamarit and Giovanni Zattera for useful conversations. FVD is supported by a Research Fellowship at Peterhouse, University of Cambridge, and would like to thank Katy Clough and Tim Dietrich for helpful discussions. This work has been partially supported by STFC consolidated grant ST/P000681/1.

	\appendix
		 \section{Boundary conditions}\label{axial}
		 We now discuss the basis solutions in the stellar interior for the frequencies of interest $\omega = \omega_{\ell n} < m \Omega$. From Maxwell's equations and the perfect conductivity condition inside the star, we have
		 \begin{equation}
		 \nabla \times \textbf{E} +\frac{\partial \textbf{B}}{\partial t}=0, \qquad \textbf{E} + \textbf{u}\times \textbf{B} =0, \qquad \text{for} \quad 0 \leq r < R,
		 \end{equation}
		 where $\textbf{u} = \boldsymbol{\Omega}\times \textbf{r}$ is the unperturbed fluid velocity in the stellar interior. This leads immediately to
		 \begin{equation}
		 \frac{\partial \textbf{B}}{\partial t}  = \nabla \times (\textbf{u} \times \textbf{B}) \label{dynamo}.
		 \end{equation}
		 For a field $\textbf{B}$ with harmonic time-dependence $\sim e^{-i \omega t}$ one finds immediately that eq. (\ref{dynamo}) leads to the following equations for the polar and radial components of $\textbf{B}$:
		 \begin{equation}
		 -i \omega  B_{r,\theta}  + \Omega \partial_\varphi B_{r,\theta} =0 .
		 \end{equation}
		 Thus for a given mode $(\ell, m)$ there are two possibilities, either $B_{r,\theta}=0$, which from the equation for $B_{\varphi}$ one finds $B_\varphi =0$, or the frequency satisfies $\omega = m \Omega$. Thus only modes with frequencies which are an integer multiple of $\Omega$ can be non-vanishing in the stellar interior, and by construction, we are interested in modes $m \Omega > \omega$. Thus for frequencies  $\omega_{\ell n} < m \Omega$ relevant for superradiance the corresponding modes vanish inside the star. Thus we must match the exterior field $A^{(i)}_{\ell}(\omega_{\ell n} r)$ to zero at $r=R$, with a finite discontinuity in $dA^{(i)}_{\ell}(\omega_{\ell n} r)/dr$. This gives a discontinuity in the associated electromagnetic fields corresponding to the generation of surface charges and currents on the surface of the perfectly conducting star.

\section{Matrix element calculations}\label{EigenCalculation}
We wish to compute the matrix elements 
\begin{align}
\bra{A^{(3)\, \ell m } } V_A \ket{ A^{(3)\, \ell m }} & = i \int_R^\infty dr\, r^2   \sigma_{_\text{M}}(r) [A^{(3)}_\ell (\omega r)]^2\left( m \Omega - \omega  \right) \label{element1},\\
\bra{   \phi_{\ell m n}  }  V_{a \gamma \gamma}(\omega_{\ell n}) \ket{  \textbf{A}^{(3)}_{\ell m}(\omega_{\ell n})} & =  - i  g_{a\gamma \gamma}\omega_{\ell n} B    
\int dr \, \,r  A^{(3)}_\ell (\omega r) \Phi_{\ell m}(r)  \label{element2},
\end{align}
which go into the perturbed eigenvalue
	\begin{align}
	&\delta \omega_{\ell m n} = \frac{ \pi^2}{8 \omega_{\ell n}}  \sum_{\ell_{1,2},m_{1,2}} \sum_{i,j} \nonumber \\
	&\bra{   \phi_{\ell m n}  }  V_{a \gamma \gamma} \ket{  A^{(i)}_{\ell_1 m_1}(\omega_{\ell n}) } 
	\bra{A^{(i)}_{\ell_1 m_1}(\omega_{\ell n})  }  V_A \ket{ A^{(j)}_{\ell_2 m_2 }(\omega_{\ell n})} 
	\bra{A^{(j)}_{\ell_2 m_2 } (\omega_{\ell n})  } V_{a \gamma \gamma} \ket{\phi_{\ell m n}}
	\end{align}
We take the solutions (\ref{Hatom}) for the axion and photon solutions. To compute the first matrix element (\ref{element1}), we use the $\omega R \ll 1$ limit to approximate:
\begin{equation}
\lim_{\omega R \ll 1}  A^{(3)}_\ell (\omega r) =  \sqrt{\frac{2 \omega}{\pi}}  j_\ell (\omega r) .
\end{equation}
The integral can then be approximated by
\begin{align}
V_{AA} = i \sigma_{_\text{M}}  \left( m \Omega - \omega  \right) \frac{2 }{\pi \omega^2}\int_{\omega R}^{\omega R_{_\text{LC}}}dx\, x^2 [j_\ell(x)]^2,
\end{align}
from which it follows that to leading order in $R\omega$ one finds:
\begin{equation}
\bra{A^{(3)\, \ell m }(\omega) } V_A \ket{ A^{(3)\, \ell m }(\omega)} \simeq i \sigma_{_\text{M}}  \left( m \Omega - \omega  \right) \frac{  (R_{ _\text{LC}} \omega)^{2\ell+3}- (R \omega)^{2\ell+3}}{ \omega^2  2^{2(\ell+1)} }\frac{1}{\left(l+\frac{3}{2}\right) \Gamma \left(l+\frac{3}{2}\right)^2}. \label{Calc1}
\end{equation}
We require the modulus squared of the second matrix element  which upon insertion of solution (\ref{Hatom}) and (\ref{Ai}) into (\ref{element2}) leads to
\begin{align}
 \bra{   \phi_{\ell m n}  }  V_{a \gamma \gamma}(\omega_{\ell n}) \ket{  \textbf{A}^{(3)}_{\ell m}(\omega_{\ell n})} = -i  g_{a\gamma \gamma} B
\sqrt{\frac{ n!}{\pi(n+ 2 \ell + 1)! (n + \ell +1)} } \left(\frac{\omega_{\ell n}}{\alpha_{\ell n}\mu} \right)^{3/2}\,I \label{Aphigam},
\end{align}
where
\begin{align}
I &= \int_{R\mu \alpha_n}^\infty dx x \,  e^{-x/2} x^{\ell+1} L_{n}^{2 \ell + 1} 
\left[ x \right]\frac{\left[ y_\ell (\omega_{\ell n} R)j_\ell\left(\frac{\omega_{\ell n}}{\mu} \frac{x}{\alpha} \right)-  j_\ell (\omega_{\ell n} R) y_\ell\left(\frac{\omega_{\ell n}}{\mu} \frac{x}{\alpha} \right)\right]}{N_\ell(\omega_{\ell n} R)}.
\end{align} 
Taking the leading order part in $\omega_{\ell n}R \lesssim 1$ gives:
\begin{equation}
\text{I} =  \int_0^\infty dx \, \Bigg[e^{-x/2}  x^{\ell+2} L_{n}^{2 \ell + 1}  
\left(   x \right)\Bigg]j_\ell\left(\frac{\omega_{\ell n}}{\mu} \frac{x}{\alpha} \right) \label{I}.
\end{equation} 
We now make use of the explicit form of the Laguerre Polynomials
\begin{equation}
L^{2 \ell + 1}_{n}(x) = \sum^n_{k=0} (-1)^k \left(
\begin{array}{c}
n+2 \ell + 1\\
n- k
\end{array}
\right) 
\frac{x^k}{k!} \label{Laguerre},
\end{equation}
   to write
   \begin{equation}
   I = \sum^n_{k=0} (-1)^k \left(
   \begin{array}{c}
   n+2 \ell + 1\\
   n- k
   \end{array}
   \right) 
   \frac{I_k}{k!} ,
   \end{equation}
   where
   \begin{align}
   I_k  & \equiv  \int_0^\infty dx \, \Bigg[e^{-x/2}  x^{\ell +k +2}  
   \Bigg]j_\ell\left(\frac{\omega_{\ell n}}{\mu} \frac{x}{\alpha} \right) \nonumber \\
   &=\frac{\sqrt{\pi } 2^{k+l+2}  \Gamma (k+2 l+3) }{\Gamma \left(l+\frac{3}{2}\right)} \left(\frac{\omega_{\ell n}}{\mu \alpha }\right)^\ell \, _2F_1\left[\frac{k}{2}+l+2,\frac{k+3}{2}+l;l+\frac{3}{2};-4 \left(\frac{\omega_{\ell n}}{\mu \alpha }\right)^2\right]. 
   \end{align}
   After expanding each $I_k$ it is easy to see that the leading $\alpha$ contribution comes from the $k=0$ term giving
   \begin{equation}
   I= \left(\frac{\mu \alpha }{\omega_{\ell n}}\right)^{\ell+4} 2^\ell (\ell+1)! \left(
   \begin{array}{c}
   n+2 \ell + 1\\
   n
   \end{array}
   \right) + \mathcal{O}(\alpha^{\ell + 5}).
   \end{equation}
   from which one finds
   \begin{align}
   & \bra{   \phi_{\ell m n}  }  V_{a \gamma \gamma}(\omega_{\ell n}) \ket{  \textbf{A}^{(3)}_{\ell m}(\omega_{\ell n})}=  -i g_{a\gamma \gamma} B \left(\frac{\mu}{\omega_{\ell n}} \right)^{\ell+5/2}\,\frac{ \alpha^{\ell +5/2}\Gamma (l+2)}{\Gamma (2 l+2)} 2^{l} \sqrt{ \frac{ \Gamma (2 l+n+2)}{\pi (l+n+1)  \Gamma (n+1)  } } \label{Calc2}.
   \end{align}

\section{Photon basis from vector spherical harmonics}\label{VecHarmonicBasis}

One can also construct a photon basis from vector spherical harmonics by considering solutions of the form
\cite{Rosa:2011my,Cardoso:2017kgn} 
\begin{equation}
A^{\ell \mu}_{\mu} = \frac{1}{r} \sum_{i=1}^4 c_i u_{i}^{\ell m}(r) Z_\mu^{(i) \, \ell m}(\theta,\varphi), \label{AAnsatz}
\end{equation}
where
\begin{align}
Z^{(1)\, \ell m }_\mu &= \left[1,0,0,0\right]Y_{\ell m} ,
\nonumber \\
Z^{(2)\, \ell m }_\mu &= \left[0,1,0,0\right]Y_{\ell m} ,
\nonumber \\
Z^{(3)\, \ell m }_\mu &=\frac{r}{\sqrt{\ell (\ell+1)}} \left[0,\, 0,\, \partial_\theta , \partial_\varphi  \right] Y_{\ell m},
\nonumber \\
Z^{(4)\, \ell m }_\mu &=\frac{r}{\sqrt{\ell (\ell+1)}} \left[0,\, 0,\,  1/\sin \theta \partial_\varphi  ,\,
-\sin \theta \partial_\theta \right] Y_{\ell m},
\end{align}
are 4-vector spherical harmonics,
and $c_1=c_2=1$, $c_{3}=c_4=[\ell(\ell+1)]^{-1/2}$. Note they satisfy the following orthonormality condition
\begin{equation}
\int d \Omega \, \,\eta^{\mu \nu}Z^{(i)\, \ell m \,  *}_\mu Z^{(j)\, \ell' m'} _\nu = \delta^{ ij} \delta_{ \ell \ell'} \delta_{m m'}. \label{Orthogonality}
\end{equation}
The modes $i = 1,2,3$ are known as polar modes, whilst $Z^{(4)}$ is the axial mode owing to their parity properties. Explicitly the vector harmonics and axion field have the following \textit{passive} parity transformation properties of their spatial components:
\begin{align}
&\phi_{\ell m n} (\textbf{x}) = (-1)^{\ell}  \phi_{\ell m n} (-\textbf{x}), \qquad \textbf{Z}^{(4)}_{\ell m} (\textbf{x}) = (-1)^{\ell} \textbf{Z}^{(4)}_{\ell m} (-\textbf{x}) ,   \nonumber\\
&\textbf{Z}^{(i)}_{\ell m} (\textbf{x}) = -(-1)^{\ell}\textbf{Z}^{(i)}_{\ell m} (-\textbf{x}), \qquad i =1,2,3 \label{Parity}.
\end{align}
Inserting the form (\ref{AAnsatz}) into (\ref{Orthogonality}) gives the following equations of motion for the radial components:
\begin{align}
\mathcal{D}^2 u_1 =0, \nonumber \\
\mathcal{D}^2 u_2 - \frac{2}{r^2}(u_2 - u_3) =0, \nonumber \\
\mathcal{D}^2 u_3 + \frac{2\ell(\ell+1)}{r^2}u_2=0, \nonumber \\
\mathcal{D}^2 u_4 =0, \label{radial}
\end{align}
where
\begin{equation}
\mathcal{D}^2 =  \omega^2 + \frac{d^2}{dr^2} - \frac{\ell(\ell+1)}{r^2}.
\end{equation}
Notice that the axial equation is completely decoupled from the polar ones, a consequence of the spherical symmetry of the photon Hamiltonian. The radial equations (\ref{radial}) are solved by a set of 4 basis solutions
\begin{equation}
\left(
\begin{array}{c}
u_1 \\
u_2 \\
u_3
\end{array}
\right) = r \left(
\begin{array}{c}
\alpha_1 i A_{\ell}(\omega r)  \\
\beta_1 \,\,A_{\ell+1}(\omega r)\\
- \beta_1 \,\, \ell \, A_{\ell+1}(\omega r) 
\end{array}
\right), \qquad \quad
r \left(
\begin{array}{c}
- \alpha_2 i A_{\ell}(\omega r)  \\
\beta_2  A_{\ell-1}(\omega r)\\
\beta_2 (\ell+1) A_{\ell-1}(\omega r)
\end{array}
\right)\qquad \alpha_i, \beta_i \in \mathbb{C} \label{basis}
\end{equation}
and $A_\ell = A_\ell(\omega r)$ is a solution to
\begin{equation}
\mathcal{D}^2 (r A_\ell) =0 . \label{Bessel}
\end{equation}
In general the $A_\ell$ will therefore be appropriately normalised linear combinations of  $j_\ell$ and $y_\ell$ - the spherical Bessel functions of the first and second kind, respectively, with the exact form being determined from the boundary conditions $\textbf{E} + \textbf{u} \times \textbf{B}=0$ for $r < R$, corresponding to infinite conductivity in the stellar interior. Using these basis solutions, we can construct a complete basis satisfying (\ref{Laplace}) and (\ref{AComplete}) given by
\begin{align}
\ket{{A^\mu_{(1)}(\omega),\ell ,m}} &=  	Z^{(1)\, \, \mu}_{\ell m}  A^{(1)}_\ell(\omega r ),  \nonumber \\
\ket{{A^\mu_{(2)}(\omega),\ell ,m}} &=  \frac{A^{(2)}_{\ell+1}(\omega r )}{(2\ell +1)^{1/2}}\left[ Z^{(2) \, \mu}_{\ell m}    - Z^{(3)\, \, \mu}_{\ell m}  c_3 \ell   \right] (\ell+1)^{1/2} ,\nonumber \\
\ket{{A^\mu_{(3)}(\omega),\ell ,m}} &= \frac{ A^{(3)}_{\ell-1}(\omega r)}{(2\ell +1)^{1/2}}\left[Z^{(2)\,\, \mu}_{\ell m}    + Z^{(3)\, \mu}_{\ell m} c_3 (\ell+1)  \right] \ell^{1/2}, \nonumber \\
\ket{{A^\mu_{(4)}(\omega),\ell ,m}} &=  A^{(4)}_\ell (\omega r)  Z^{(4) \, \mu}_{\ell m}.\label{Basis}
\end{align}
where the superscripts on the $s^{(i)}_\ell$ remind us that they may be solutions to the order $\ell$ spherical Bessel's equation with different boundary conditions. The basis is clearly complete since
\begin{equation}
\sum_{\ell \, m} \sum_{i=1}^4 \, \,Z^{(i)\, \ell m}_\mu (\hat{\textbf{x}}) Z_\nu^{(i) \, \ell m \, \,  *}(\hat{\textbf{x}}') = \eta_{\mu \nu} \delta(\hat{\textbf{x}} ,\hat{\textbf{x}}'). \label{ZComplete}
\end{equation} 
and
\begin{equation}
\int d[\omega^2] A_{\ell}(\omega r) A_\ell(\omega r') = \frac{1}{r^2}\delta(r-r'),
\end{equation}
which follows by virtue of the Sturm-Liouville nature of Bessel's equation.
	\bibliography{References_Axion_Configurations}{}

\providecommand{\href}[2]{#2}\begingroup\raggedright\begin{thebibliography}{10}

\bibitem{Kokkotas:1999bd}
K.~D. Kokkotas and B.~G. Schmidt, \emph{{Quasinormal modes of stars and black
  holes}}, \href{https://doi.org/10.12942/lrr-1999-2}{\emph{Living Rev. Rel.}
  {\bfseries 2} (1999) 2},
  [\href{https://arxiv.org/abs/gr-qc/9909058}{{\ttfamily gr-qc/9909058}}].

\bibitem{Konoplya:2011qq}
R.~A. Konoplya and A.~Zhidenko, \emph{{Quasinormal modes of black holes: From
  astrophysics to string theory}},
  \href{https://doi.org/10.1103/RevModPhys.83.793}{\emph{Rev. Mod. Phys.}
  {\bfseries 83} (2011) 793--836},
  [\href{https://arxiv.org/abs/1102.4014}{{\ttfamily 1102.4014}}].

\bibitem{Pani:2013pma}
P.~Pani, \emph{{Advanced Methods in Black-Hole Perturbation Theory}},
  \href{https://doi.org/10.1142/S0217751X13400186}{\emph{Int. J. Mod. Phys.}
  {\bfseries A28} (2013) 1340018},
  [\href{https://arxiv.org/abs/1305.6759}{{\ttfamily 1305.6759}}].

\bibitem{Andersson:2000mf}
N.~Andersson and K.~D. Kokkotas, \emph{{The R mode instability in rotating
  neutron stars}}, \href{https://doi.org/10.1142/S0218271801001062}{\emph{Int.
  J. Mod. Phys.} {\bfseries D10} (2001) 381--442},
  [\href{https://arxiv.org/abs/gr-qc/0010102}{{\ttfamily gr-qc/0010102}}].

\bibitem{Svrcek:2006yi}
P.~Svrcek and E.~Witten, \emph{{Axions In String Theory}},
  \href{https://doi.org/10.1088/1126-6708/2006/06/051}{\emph{JHEP} {\bfseries
  06} (2006) 051}, [\href{https://arxiv.org/abs/hep-th/0605206}{{\ttfamily
  hep-th/0605206}}].

\bibitem{Arvanitaki:2009fg}
A.~Arvanitaki, S.~Dimopoulos, S.~Dubovsky, N.~Kaloper and J.~March-Russell,
  \emph{{String Axiverse}},
  \href{https://doi.org/10.1103/PhysRevD.81.123530}{\emph{Phys. Rev.}
  {\bfseries D81} (2010) 123530},
  [\href{https://arxiv.org/abs/0905.4720}{{\ttfamily 0905.4720}}].

\bibitem{Arvanitaki:2010sy}
A.~Arvanitaki and S.~Dubovsky, \emph{{Exploring the String Axiverse with
  Precision Black Hole Physics}},
  \href{https://doi.org/10.1103/PhysRevD.83.044026}{\emph{Phys. Rev.}
  {\bfseries D83} (2011) 044026},
  [\href{https://arxiv.org/abs/1004.3558}{{\ttfamily 1004.3558}}].

\bibitem{Pani:2012vp}
P.~Pani, V.~Cardoso, L.~Gualtieri, E.~Berti and A.~Ishibashi, \emph{{Black hole
  bombs and photon mass bounds}},
  \href{https://doi.org/10.1103/PhysRevLett.109.131102}{\emph{Phys. Rev. Lett.}
  {\bfseries 109} (2012) 131102},
  [\href{https://arxiv.org/abs/1209.0465}{{\ttfamily 1209.0465}}].

\bibitem{Arvanitaki:2014wva}
A.~Arvanitaki, M.~Baryakhtar and X.~Huang, \emph{{Discovering the QCD Axion
  with Black Holes and Gravitational Waves}},
  \href{https://doi.org/10.1103/PhysRevD.91.084011}{\emph{Phys. Rev.}
  {\bfseries D91} (2015) 084011},
  [\href{https://arxiv.org/abs/1411.2263}{{\ttfamily 1411.2263}}].

\bibitem{Brito:2014wla}
R.~Brito, V.~Cardoso and P.~Pani, \emph{{Black holes as particle detectors:
  evolution of superradiant instabilities}},
  \href{https://doi.org/10.1088/0264-9381/32/13/134001}{\emph{Class. Quant.
  Grav.} {\bfseries 32} (2015) 134001},
  [\href{https://arxiv.org/abs/1411.0686}{{\ttfamily 1411.0686}}].

\bibitem{Brito:2015oca}
R.~Brito, V.~Cardoso and P.~Pani, \emph{{Superradiance}},
  \href{https://doi.org/10.1007/978-3-319-19000-6}{\emph{Lect. Notes Phys.}
  {\bfseries 906} (2015) pp.1--237},
  [\href{https://arxiv.org/abs/1501.06570}{{\ttfamily 1501.06570}}].

\bibitem{Rosa:2017ury}
J.~G. Rosa and T.~W. Kephart, \emph{{Black hole lasers powered by axion
  superradiant instabilities}},
  \href{https://arxiv.org/abs/1709.06581}{{\ttfamily 1709.06581}}.

\bibitem{Filippini:2019cqk}
F.~Filippini and G.~Tasinato, \emph{{On long range axion hairs for black
  holes}},  \href{https://arxiv.org/abs/1903.02950}{{\ttfamily 1903.02950}}.

\bibitem{Cardoso:2015zqa}
V.~Cardoso, R.~Brito and J.~L. Rosa, \emph{{Superradiance in stars}},
  \href{https://doi.org/10.1103/PhysRevD.91.124026}{\emph{Phys. Rev.}
  {\bfseries D91} (2015) 124026},
  [\href{https://arxiv.org/abs/1505.05509}{{\ttfamily 1505.05509}}].

\bibitem{Richartz:2013hza}
M.~Richartz and A.~Saa, \emph{{Superradiance without event horizons in General
  Relativity}}, \href{https://doi.org/10.1103/PhysRevD.88.044008}{\emph{Phys.
  Rev.} {\bfseries D88} (2013) 044008},
  [\href{https://arxiv.org/abs/1306.3137}{{\ttfamily 1306.3137}}].

\bibitem{Cardoso:2017kgn}
V.~Cardoso, P.~Pani and T.-T. Yu, \emph{{Superradiance in rotating stars and
  pulsar-timing constraints on dark photons}},
  \href{https://doi.org/10.1103/PhysRevD.95.124056}{\emph{Phys. Rev.}
  {\bfseries D95} (2017) 124056},
  [\href{https://arxiv.org/abs/1704.06151}{{\ttfamily 1704.06151}}].

\bibitem{ATNF}
R.~N. Manchester, G.~B. Hobbs, A.~Teoh and M.~Hobbs, \emph{{The Australia
  Telescope National Facility pulsar catalogue}},
  \href{https://doi.org/10.1086/428488}{\emph{Astron. J.} {\bfseries 129}
  (2005) 1993}, [\href{https://arxiv.org/abs/astro-ph/0412641}{{\ttfamily
  astro-ph/0412641}}].

\bibitem{Lorimer}
D.~R. {Lorimer} and M.~{Kramer}, \emph{{Handbook of Pulsar Astronomy}}.
\newblock Oct., 2012.

\bibitem{Plascencia:2017kca}
A.~D. Plascencia and A.~Urbano, \emph{{Black hole superradiance and
  polarization-dependent bending of light}},
  \href{https://doi.org/10.1088/1475-7516/2018/04/059}{\emph{JCAP} {\bfseries
  1804} (2018) 059}, [\href{https://arxiv.org/abs/1711.08298}{{\ttfamily
  1711.08298}}].

\bibitem{Mohanty:1993nh}
S.~Mohanty and S.~N. Nayak, \emph{{Determination of pseudoGoldstone boson -
  photon coupling by the differential time delay of pulsar signals}},
  \href{https://doi.org/10.1103/PhysRevLett.70.4038,
  10.1103/PhysRevLett.71.1117, 10.1103/PhysRevLett.76.2825}{\emph{Phys. Rev.
  Lett.} {\bfseries 70} (1993) 4038--4041},
  [\href{https://arxiv.org/abs/hep-ph/9303310}{{\ttfamily hep-ph/9303310}}].

\bibitem{1510.07633}
D.~J.~E. Marsh, \emph{{Axion Cosmology}},
  \href{https://doi.org/10.1016/j.physrep.2016.06.005}{\emph{Phys. Rept.}
  {\bfseries 643} (2016) 1--79},
  [\href{https://arxiv.org/abs/1510.07633}{{\ttfamily 1510.07633}}].

\bibitem{Peccei:1977hh}
R.~D. Peccei and H.~R. Quinn, \emph{{CP Conservation in the Presence of
  Instantons}}, \href{https://doi.org/10.1103/PhysRevLett.38.1440}{\emph{Phys.
  Rev. Lett.} {\bfseries 38} (1977) 1440--1443}.

\bibitem{Weinberg:1977ma}
S.~Weinberg, \emph{{A New Light Boson?}},
  \href{https://doi.org/10.1103/PhysRevLett.40.223}{\emph{Phys. Rev. Lett.}
  {\bfseries 40} (1978) 223--226}.

\bibitem{Wilczek:1977pj}
F.~Wilczek, \emph{{Problem of Strong $P$ and $T$ Invariance in the Presence of
  Instantons}}, \href{https://doi.org/10.1103/PhysRevLett.40.279}{\emph{Phys.
  Rev. Lett.} {\bfseries 40} (1978) 279--282}.

\bibitem{hep-th/0602233}
J.~P. Conlon, \emph{{The QCD axion and moduli stabilisation}},
  \href{https://doi.org/10.1088/1126-6708/2006/05/078}{\emph{JHEP} {\bfseries
  05} (2006) 078}, [\href{https://arxiv.org/abs/hep-th/0602233}{{\ttfamily
  hep-th/0602233}}].

\bibitem{hep-th/0605206}
P.~Svrcek and E.~Witten, \emph{{Axions In String Theory}},
  \href{https://doi.org/10.1088/1126-6708/2006/06/051}{\emph{JHEP} {\bfseries
  06} (2006) 051}, [\href{https://arxiv.org/abs/hep-th/0605206}{{\ttfamily
  hep-th/0605206}}].

\bibitem{Patrignani:2016xqp}
{\scshape Particle Data Group} collaboration, C.~Patrignani et~al.,
  \emph{{Review of Particle Physics}},
  \href{https://doi.org/10.1088/1674-1137/40/10/100001}{\emph{Chin. Phys.}
  {\bfseries C40} (2016) 100001}.

\bibitem{Payez:2014xsa}
A.~Payez, C.~Evoli, T.~Fischer, M.~Giannotti, A.~Mirizzi and A.~Ringwald,
  \emph{{Revisiting the SN1987A gamma-ray limit on ultralight axion-like
  particles}}, \href{https://doi.org/10.1088/1475-7516/2015/02/006}{\emph{JCAP}
  {\bfseries 1502} (2015) 006},
  [\href{https://arxiv.org/abs/1410.3747}{{\ttfamily 1410.3747}}].

\bibitem{Berg:2016ese}
M.~Berg, J.~P. Conlon, F.~Day, N.~Jennings, S.~Krippendorf, A.~J. Powell
  et~al., \emph{{Constraints on Axion-Like Particles from X-ray Observations of
  NGC1275}}, \href{https://doi.org/10.3847/1538-4357/aa8b16}{\emph{Astrophys.
  J.} {\bfseries 847} (2017) 101},
  [\href{https://arxiv.org/abs/1605.01043}{{\ttfamily 1605.01043}}].

\bibitem{Marsh:2017yvc}
M.~C.~D. Marsh, H.~R. Russell, A.~C. Fabian, B.~P. McNamara, P.~Nulsen and
  C.~S. Reynolds, \emph{{A New Bound on Axion-Like Particles}},
  \href{https://doi.org/10.1088/1475-7516/2017/12/036}{\emph{JCAP} {\bfseries
  1712} (2017) 036}, [\href{https://arxiv.org/abs/1703.07354}{{\ttfamily
  1703.07354}}].

\bibitem{Conlon:2017qcw}
J.~P. Conlon, F.~Day, N.~Jennings, S.~Krippendorf and M.~Rummel,
  \emph{{Constraints on Axion-Like Particles from Non-Observation of Spectral
  Modulations for X-ray Point Sources}},
  \href{https://doi.org/10.1088/1475-7516/2017/07/005}{\emph{JCAP} {\bfseries
  1707} (2017) 005}, [\href{https://arxiv.org/abs/1704.05256}{{\ttfamily
  1704.05256}}].

\bibitem{Garbrecht:2018akc}
B.~Garbrecht and J.~I. McDonald, \emph{{Axion configurations around pulsars}},
  \href{https://doi.org/10.1088/1475-7516/2018/07/044}{\emph{JCAP} {\bfseries
  1807} (2018) 044}, [\href{https://arxiv.org/abs/1804.04224}{{\ttfamily
  1804.04224}}].

\bibitem{1811.04945}
M.~Boskovic, R.~Brito, V.~Cardoso, T.~Ikeda and H.~Witek, \emph{{Axionic
  instabilities and new black hole solutions}},
  \href{https://arxiv.org/abs/1811.04945}{{\ttfamily 1811.04945}}.

\bibitem{LANDAU1977133}
L.~LANDAU and E.~LIFSHITZ, \emph{Chapter vi - perturbation theory},  in
  \emph{Quantum Mechanics (Third Edition)} (L.~LANDAU and E.~LIFSHITZ, eds.),
  pp.~133 -- 163.
\newblock Pergamon, third edition~ed., 1977.
\newblock
  \href{https://doi.org/https://doi.org/10.1016/B978-0-08-020940-1.50013-X}{DOI}.

\bibitem{Cohen:2006fh}
D.~Cohen, \emph{{Lecture notes in quantum mechanics}},
  \href{https://arxiv.org/abs/quant-ph/0605180}{{\ttfamily quant-ph/0605180}}.

\bibitem{Baiko:1995qg}
D.~A. Baiko and D.~G. Yakovlev, \emph{{Thermal and electric conductivities of
  Coulomb crystals in neutron stars and white dwarfs}}, {\emph{Astron. Lett.}
  {\bfseries 21} (1995) 709},
  [\href{https://arxiv.org/abs/astro-ph/9604164}{{\ttfamily
  astro-ph/9604164}}].

\bibitem{10.2307/43638547}
R.~K.~M. THAMBYNAYAGAM and T.~M. HABASHY, \emph{A new weber-type transform},
  {\emph{Quarterly of Applied Mathematics} {\bfseries 61} (2003) 485--493}.

\bibitem{Hessels:2006ze}
J.~W.~T. Hessels, S.~M. Ransom, I.~H. Stairs, P.~C.~C. Freire, V.~M. Kaspi and
  F.~Camilo, \emph{{A radio pulsar spinning at 716-hz}},
  \href{https://doi.org/10.1126/science.1123430}{\emph{Science} {\bfseries 311}
  (2006) 1901--1904}, [\href{https://arxiv.org/abs/astro-ph/0601337}{{\ttfamily
  astro-ph/0601337}}].

\bibitem{Harding:2013ij}
A.~K. Harding, \emph{{The Neutron Star Zoo}},
  \href{https://doi.org/10.1007/s11467-013-0285-0}{\emph{Front. Phys.(Beijing)}
  {\bfseries 8} (2013) 679--692},
  [\href{https://arxiv.org/abs/1302.0869}{{\ttfamily 1302.0869}}].

\bibitem{Ikeda:2019fvj}
T.~Ikeda, R.~Brito and V.~Cardoso, \emph{{Blasts of Light from Axions}},
  \href{https://doi.org/10.1103/PhysRevLett.122.081101}{\emph{Phys. Rev. Lett.}
  {\bfseries 122} (2019) 081101},
  [\href{https://arxiv.org/abs/1811.04950}{{\ttfamily 1811.04950}}].

\bibitem{Tercas:2018gxv}
H.~Terças, J.~D. Rodrigues and J.~T. Mendonça, \emph{{Axion-plasmon
  polaritons in strongly magnetized plasmas}},
  \href{https://doi.org/10.1103/PhysRevLett.120.181803}{\emph{Phys. Rev. Lett.}
  {\bfseries 120} (2018) 181803},
  [\href{https://arxiv.org/abs/1801.06254}{{\ttfamily 1801.06254}}].

\bibitem{Mendonca:2019eke}
J.~T. Mendonça, J.~D. Rodrigues and H.~Terças, \emph{{Axion production in
  unstable magnetized plasmas: an active source of dark-matter}},
  \href{https://arxiv.org/abs/1901.05910}{{\ttfamily 1901.05910}}.

\bibitem{McDonald}
J.~I. McDonald, \emph{{to appear}}, .

\bibitem{Li:2011zh}
J.~Li, A.~Spitkovsky and A.~Tchekhovskoy, \emph{{Resistive Solutions for Pulsar
  Magnetospheres}},
  \href{https://doi.org/10.1088/0004-637X/746/1/60}{\emph{Astrophys. J.}
  {\bfseries 746} (2012) 60},
  [\href{https://arxiv.org/abs/1107.0979}{{\ttfamily 1107.0979}}].

\bibitem{Kalapotharakos:2011vg}
C.~Kalapotharakos, D.~Kazanas, A.~Harding and I.~Contopoulos, \emph{{Toward a
  Realistic Pulsar Magnetosphere}},
  \href{https://doi.org/10.1088/0004-637X/749/1/2}{\emph{Astrophys. J.}
  {\bfseries 749} (2012) 2}, [\href{https://arxiv.org/abs/1108.2138}{{\ttfamily
  1108.2138}}].

\bibitem{Kalapotharakos:2013sma}
C.~Kalapotharakos, A.~K. Harding and D.~Kazanas, \emph{{Gamma-Ray Emission in
  Dissipative Pulsar Magnetospheres: From Theory to Fermi Observations}},
  \href{https://doi.org/10.1088/0004-637X/793/2/97}{\emph{Astrophys. J.}
  {\bfseries 793} (2014) 97},
  [\href{https://arxiv.org/abs/1310.3545}{{\ttfamily 1310.3545}}].

\bibitem{Brambilla:2015vta}
G.~Brambilla, C.~Kalapotharakos, A.~K. Harding and D.~Kazanas, \emph{{Testing
  dissipative magnetosphere model light curves and spectra with FERMI
  pulsars}}, \href{https://doi.org/10.1088/0004-637X/804/2/84}{\emph{Astrophys.
  J.} {\bfseries 804} (2015) 84},
  [\href{https://arxiv.org/abs/1503.00744}{{\ttfamily 1503.00744}}].

\bibitem{Conlon:2017hhi}
J.~P. Conlon and C.~A.~R. Herdeiro, \emph{{Can black hole superradiance be
  induced by galactic plasmas?}},
  \href{https://doi.org/10.1016/j.physletb.2018.02.073}{\emph{Phys. Lett.}
  {\bfseries B780} (2018) 169--173},
  [\href{https://arxiv.org/abs/1701.02034}{{\ttfamily 1701.02034}}].

\bibitem{Sen:2018cjt}
S.~Sen, \emph{{Plasma effects on lasing of a uniform ultralight axion
  condensate}}, \href{https://doi.org/10.1103/PhysRevD.98.103012}{\emph{Phys.
  Rev.} {\bfseries D98} (2018) 103012},
  [\href{https://arxiv.org/abs/1805.06471}{{\ttfamily 1805.06471}}].

\bibitem{Kalapotharakos:2017tzm}
C.~Kalapotharakos, A.~K. Harding, D.~Kazanas and G.~Brambilla, \emph{{Fermi
  Gamma-Ray Pulsars: Understanding the High-Energy Emission from Dissipative
  Magnetospheres}},
  \href{https://doi.org/10.3847/1538-4357/aa713a}{\emph{Astrophys. J.}
  {\bfseries 842} (2017) 80},
  [\href{https://arxiv.org/abs/1702.03069}{{\ttfamily 1702.03069}}].

\bibitem{PhysRev.70.202}
W.~M. Elsasser, \emph{Induction effects in terrestrial magnetism part ii. the
  secular variation}, \href{https://doi.org/10.1103/PhysRev.70.202}{\emph{Phys.
  Rev.} {\bfseries 70} (Aug, 1946) 202--212}.

\bibitem{2007egp..book.....G}
D.~{Gubbins} and E.~{Herrero-Bervera}, \emph{{Encyclopedia of Geomagnetism and
  Paleomagnetism}}.
\newblock 2007.

\bibitem{Melrose:2016kaf}
D.~B. Melrose and R.~Yuen, \emph{{Pulsar Electrodynamics: an unsolved
  problem}}, \href{https://doi.org/10.1017/S0022377816000398}{\emph{J. Plasma
  Phys.} {\bfseries 82} (2016) 635820202},
  [\href{https://arxiv.org/abs/1604.03623}{{\ttfamily 1604.03623}}].

\bibitem{Deutsch}
A.~J. {Deutsch}, \emph{{The electromagnetic field of an idealized star in rigid
  rotation in vacuo}}, {\emph{Annales d'Astrophysique} {\bfseries 18} (Jan.,
  1955) 1}.

\bibitem{Rosa:2011my}
J.~G. Rosa and S.~R. Dolan, \emph{{Massive vector fields on the Schwarzschild
  spacetime: quasi-normal modes and bound states}},
  \href{https://doi.org/10.1103/PhysRevD.85.044043}{\emph{Phys. Rev.}
  {\bfseries D85} (2012) 044043},
  [\href{https://arxiv.org/abs/1110.4494}{{\ttfamily 1110.4494}}].

\end{thebibliography}\endgroup
	\bibliographystyle{JHEP}

\end{document}